\documentclass[12pt]{article}
\usepackage{amsmath,amssymb,amsbsy,amstext,amsthm,booktabs,simplewick,exscale,relsize,slashed,graphicx,amsfonts,upgreek}
\usepackage{multirow,dcolumn,bm,enumerate,url,hyperref,cite}
\usepackage{color}

\usepackage[T1]{fontenc} 
\usepackage{subcaption}
\usepackage[capitalize]{cleveref}

\newcommand{\mAp}{m_{A'}}
\newcommand{\calM}{\mathcal{M}}
\newcommand{\ph}{\mathrm{pH}_2}
\newcommand{\e}{|e\rangle}
\newcommand{\g}{|g\rangle}
\newcommand{\wg}{\omega_g}
\newcommand{\we}{\omega_e}
\newcommand{\wj}{\omega_j}
\newcommand{\wjg}{\omega_{jg}}
\newcommand{\wje}{\omega_{je}}
\newcommand{\weg}{\omega_{eg}}
\newcommand{\cg}{c_g}
\newcommand{\ce}{c_e}
\newcommand{\cjp}{c_{j+}}
\newcommand{\cjm}{c_{j-}}
\newcommand{\djg}{d_{jg}}
\newcommand{\dje}{d_{je}}
\newcommand{\djgp}{d'_{jg}}
\newcommand{\djep}{d'_{je}}
\newcommand{\dgj}{d_{gj}}
\newcommand{\dej}{d_{ej}}
\newcommand{\dgjp}{d'_{gj}}
\newcommand{\dejp}{d'_{ej}}
\newcommand{\Eob}{\bar{E}_1}
\newcommand{\Etb}{\bar{E}_2}
\newcommand{\Epb}{\bar{E}'}
\newcommand{\Wee}{\Omega_{ee}}
\newcommand{\Wgg}{\Omega_{gg}}
\newcommand{\Weg}{\Omega_{eg}}
\newcommand{\Wge}{\Omega_{ge}}
\newcommand{\ree}{\rho_{ee}}
\newcommand{\rgg}{\rho_{gg}}
\newcommand{\reg}{\rho_{eg}}
\newcommand{\rge}{\rho_{ge}}
\newcommand{\aee}{a_{ee}}
\newcommand{\agg}{a_{gg}}
\newcommand{\age}{a_{ge}}
\newcommand{\aeg}{a_{eg}}
\newcommand{\tP}{\tilde{P}}

\DeclareMathOperator{\Tr}{Tr}
\DeclareMathOperator{\im}{Im}
\DeclareMathOperator{\re}{Re}
\renewcommand\vec{\mathbf}

\title{Superradiant Searches for Dark Photons in Two Stage Atomic Transitions}

\author{Amit Bhoonah$^*$, Joseph Bramante$^{*, \dagger}$, and Ningqiang Song$^{*, \dagger}$\\
\small $^*$The Arthur B. McDonald Canadian Astroparticle Physics Research Institute, \\ \small Department of Physics, Engineering Physics, and Astronomy,\\ \small Queen's University, Kingston, Ontario, K7L 2S8, Canada\\
\small $^\dagger$Perimeter Institute for Theoretical Physics, Waterloo, Ontario, N2L 2Y5, Canada}
\date{}

\begin{document}
\maketitle
\begin{abstract}
We study a new mechanism to discover dark photon fields, by resonantly triggering two photon transitions in cold gas preparations. Using coherently prepared cold parahydrogen, coupling sensitivity for sub-meV mass dark photon fields can be advanced by orders of magnitude, with a modified light-shining-through-wall setup. We calculate the effect of a background dark photon field on the dipole moment and corresponding transition rate of cold parahydrogen pumped into its first vibrational excited state by counter-propagating laser beams. The nonlinear amplification of two photon emission triggered by dark photons in a cold parahydrogen sample is numerically simulated to obtain the expected dark photon coupling sensitivity.
\end{abstract}

\newpage
\tableofcontents

\section{Introduction}
\label{sec:intro}
The first glimmer of physics beyond the Standard Model may come from a new massive U(1) gauge boson weakly mixed with the photon, sometimes called a dark photon \cite{Holdom:1985ag,delAguila:1988jz,delAguila:1995rb}. Dark photons are a predicted feature in supersymmetric theories, string theories, and hidden portal models of dark matter \cite{Dienes:1996zr,Abel:2008ai,Goodsell:2009xc,Pospelov:2008zw,Ackerman:mha,ArkaniHamed:2008qn}. Many searches are underway to detect dark photons, either produced by a star \cite{An:2013yfc,An:2013yua,Redondo:2013lna,Vinyoles:2015aba,Chang:2016ntp,Hardy:2016kme}, a laser \cite{Ehret:2010mh,Bahre:2013ywa,Pugnat:2007nu}, at colliders \cite{Kumar:2006gm,Feldman:2006wb,Bramante:2011qc,Curtin:2014cca}, or produced in the primordial universe \cite{Horns:2012jf,Jaeckel:2013eha,TheMADMAXWorkingGroup:2016hpc,Arvanitaki:2017nhi,Baryakhtar:2018doz}. If they are produced in the early universe, sub-MeV mass dark photons are a candidate model for dark matter \cite{1983PhLB..122..221B,Nelson:2011sf,Arias:2012az,Graham:2015rva,Dror:2018pdh,Agrawal:2018vin,Co:2018lka,Long:2019lwl}.
\\

Using two stage atomic transitions, this paper proposes a new method to enhance the detection of dark photons produced by lasers shining through walls. This proposal involves a dark photon field produced in a laser cavity, then passed through a sample of quasi-stable coherently excited atoms whose $E1$ dipole transitions are parity-forbidden. Under these conditions, during the brief time that the excited atoms are coherent, the diminutive field of the dark photon can resonantly trigger two-photon electronic transitions. As compared to traditional light-shining-through-wall experiments, we project a large gain in sensitivity to dark photons with $\mu$eV-meV masses. These sensitivity gains appear within reach using preparations of parahydrogen (pH$_2$) coherently excited by counter-propagating nanosecond laser pulses \cite{Miyamoto:2015tva,Hiraki:2018jwu}.
\\

The use of two stage superradiant atomic transitions for the production and detection of weakly coupled particles was proposed and studied extensively by Yoshimura et al. \cite{Yoshimura:2006nd,Yoshimura:2008ya,Fukumi:2012rn,Yoshimura:2012tm,Miyamoto:2014,Miyamoto:2015tva,Hara:2017,2017JPCA..121.3943M,Hiraki:2018jwu}. These authors along with \cite{Song:2015xaa,Boyero:2015eqa,Vaquero:2016ovj,Zhang:2016lqp} have studied how macroscopic quantities of coherently excited atoms may be employed to measure neutrino properties. The use of atomic transitions for the discovery of axion and dark photon dark matter has also recently been considered in \cite{Yoshimura:2017ghb,Huang:2019rmc,Alvarez-Luna:2018jsb,Flambaum:2019cqi,Sikivie:2014lha}. In contrast, the experiment we propose is sensitive to any U(1) vector bosons kinetically-mixed with the Standard Model photon, whether or not dark matter is comprised of a dark photon.
\\

Coherent superradiant emission by atomic systems was formalized by Dicke in \cite{Dicke:1954zz}. However, the possibility that superradiance might be observed in macroscopic amalgams of material has received increased attention in the last decade after being proposed as a method to measure certain neutrino properties \cite{Yoshimura:2008ya,Fukumi:2012rn}. Before proceeding further, we will develop some physical intuition about classic (aka Dicke) versus macro (aka Yoshimura) superradiance. A formal derivation can be found in Appendix \ref{sec:tpeqed}. Let us consider a group of atomic emitters with number density $n$ occupying volume $V$, which have been prepared in excited states, such that each excited atomic emitter is indistinguishable from the next (we want them to have the same phase). Let us suppose that some atom in volume $V$ de-excites and emits a photon with momentum $\vec k_1$. Then if a single, isolated atom has a photon emission rate $\Gamma_0$, and for the moment neglecting superradiant effects (superradiant effects would indeed be negligible if the spacing between atomic emitters is much greater than wavelength of photons emitted, $ n^{-1/3} \gg k_1^{-1}$) the emission rate of photons from volume $V$ follows trivially, $\Gamma_{tot} = nV \Gamma_0$.
\\

\begin{figure}[!t]
\centering
\includegraphics[width=0.95\textwidth]{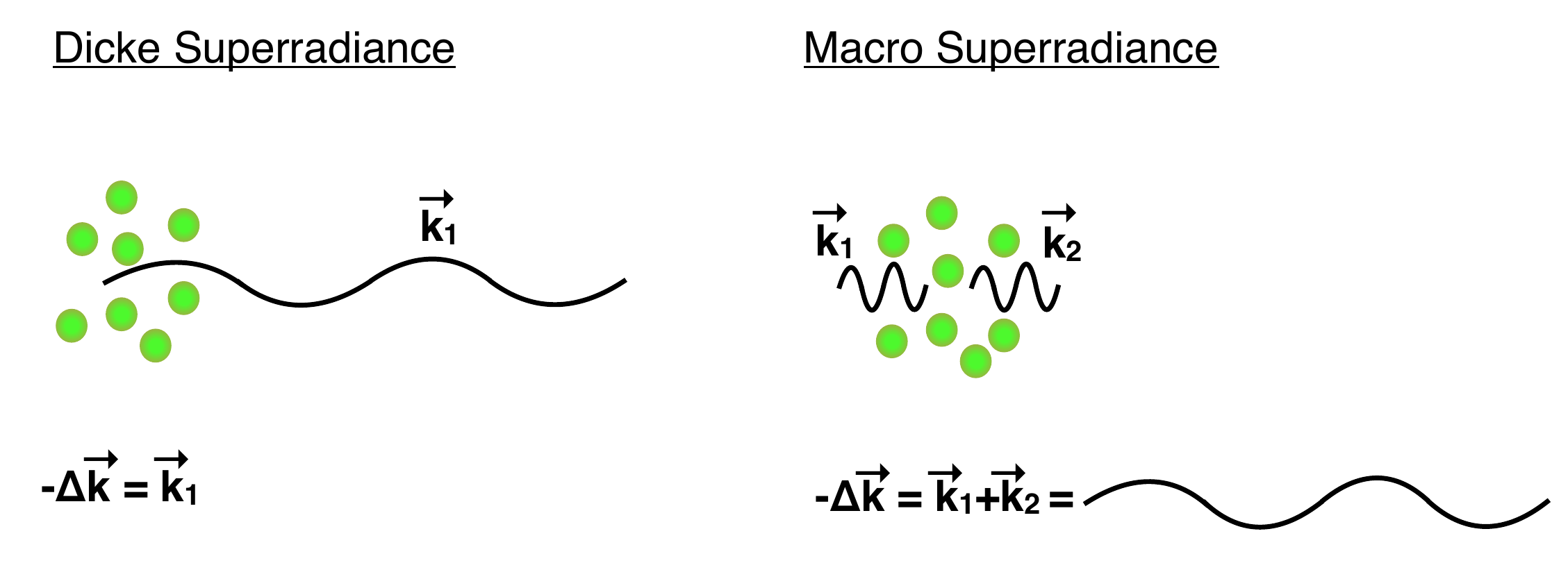}
\caption{Illustration of single photon emission, aka Dicke superradiance and two photon emission, aka Yoshimura superradiance. Crucially, the volume for superradiant emission is determined by the final state momentum $\Delta k$ of the atomic emitters. For classic Dicke superradiance, this is just the momentum of the emitted photons $\Delta k = | \vec k_1|$. For two photons emitted back-to-back, the final state momentum can be tiny, \mbox{$\Delta k \approx | \vec k_1 - \vec k_2 | \rightarrow 0$}. Of course, superradiant emission will also depend on the linewidth of lasers used to excite the atoms, material dephasing effects, and other factors, see Section \ref{sec:2state}.} \label{fig:s2r}
\end{figure}

However, if the wavelength of the emitted photon is larger than the inter-atomic spacing, there will be a superradiant enhancement to the rate of photon emission. In fact, in the case that the emitted photon's wavelength is much greater than the volume itself ($k_1^{-3} \gg V$) the total rate of photon emission will be $\Gamma_{tot} = n^2V^2 \Gamma_0$, because of superradiance. This superradiant enhancement can be understood from basic quantum principles as follows. Firstly, momentum conservation tells us that an emitting atom in its final state will have a momentum of size $\sim  k_1$. The uncertainty principle tells us that the atom is only localized over a distance $\sim \frac{1}{ k_1}$. Altogether these imply that it is not possible to determine which atom in the volume $k_1^{-3}$ emitted the photon. But we know that quantum mechanics tells us that the probability for an event to occur is the squared sum of all ways for the event to occur, and if we cannot distinguish between atoms, then we must sum over all atoms in the coherence volume $k_1^{-3}$, and square that sum to obtain the probability for emission. From this we obtain an extra factor of $nV$ in our superradiantly enhanced emission rate. This is illustrated in Figure \ref{fig:s2r}.
\\

In the preceding heuristic argument, it is crucial to realize that it is the final state momentum and corresponding spatial uncertainty of the emitting atoms that determines the volume over which superradiant emission can occur. Therefore, a process that somehow reduces the final state momentum of atomic emitters, can potentially result in superradiant emission over volumes larger than the wavelength of emitted photons. Perhaps the simplest such process is two photon emission. In two photon emission, the final state momentum of the emitting atom will be the sum of emitted momenta $-\Delta \vec k = \vec  k_1 + \vec k_2$. For back-to-back two photon emission where $\vec k_1 \approx - \vec k_2 $, the superradiant emission volume $\Delta k^{-3}$ can in principle be arbitrarily large -- it is only limited by the difference in momentum of the emitted photons. This is illustrated in Figure \ref{fig:s2r}. In practice, dephasing of atoms, the related decoherence time of the atomic medium, and the linewidth of lasers used to excited the atoms will also limit the superradiant emission volume. It is interesting to note that much of the preceding logic about cooperative emission of photons could be equally applied to cooperative absorption.
\\

In this paper we show how two photon superradiant emission by a large number of atoms can be used to detect dark photons. First, a sample of cold parahydrogen (or another suitable atomic target) is excited into a metastable state by back-to-back photons provided by counter-propagating lasers with frequency $\omega_1$. This excited state will preferentially decay through two photon emission, in part because of the macroscopic superradiant enhancement detailed above. Next, dark photons produced in an adjacent cavity by a laser with frequency $\omega_1$ are passed through the excited parahydrogen sample. The dark photon field through its mixing with the visible photon, would act as a ``trigger laser,'' resonantly de-exciting the excited parahydrogen through a two-photon transition. This altogether provides a new, very sensitive method to search for dark photons.
\\

The rest of this paper proceeds as follows. In Section \ref{sec:2state}, we calculate coherence in two stage electronic transitions, and study how much coherence is attained in cold preparations of parahydrogen excited by counter-propagating lasers. We find that the coherence necessary to begin realizing this proposal has been obtained in a number of experiments. The interaction of a dark photon with coherently excited two stage atomic systems is derived in Section \ref{sec:darkSR}. Enthusiastic readers may wish to skip to Section \ref{sec:sens}, which includes a schematic description of the experiment, along with its sensitivity to a kinetically mixed dark photon. Determining this dark photon sensitivity requires numerically integrating the dark and visible photon field equations in a background of coherently excited atoms. Conclusions are presented in Section \ref{sec:conc}. Throughout we use natural units where $\hbar =c = k_B=1$.

\section{Coherence in two stage atomic transitions}
\label{sec:2state}

Pulses from high-power lasers allow for the preparation of atoms in coherent excited states, from which they can be cooperatively de-excited. Before investigating how the weak electromagnetic field sourced by a dark photon can be detected by cooperatively de-exciting coherently prepared atoms, it will be useful to examine under what conditions counter-propagating lasers excite highly coherent atoms in the first place. After deriving the coherence of atomic states excited by counter-propagating pairs of photons, we will examine how laser power, atomic density, and temperature alter this coherence. The derivation given below can be found in many prior references \cite{Narducci:1977,Harris:97}. Our aim here is to quantify the experimental capability, in terms of coherently excited atoms, that will be needed to detect dark photons.

\subsection{Quasi-stable excited states}

\begin{figure}[!t]
\centering
\includegraphics[width=0.4\textwidth]{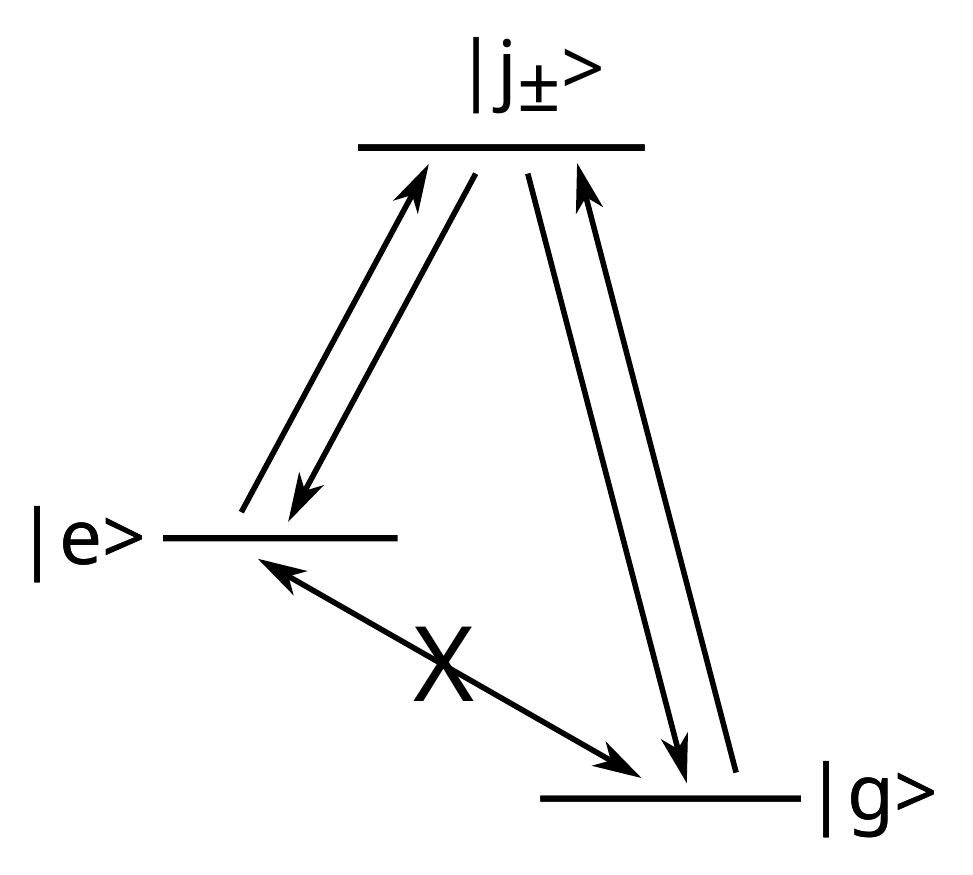}
\caption{Illustration of the energy levels of an atomic system, with ground state $| g \rangle$ and excited state $| e \rangle$. $E1$ dipole transitions between $| g \rangle$ and $| e \rangle$ are forbidden; the two-step process of transtioning from  $| g \rangle$ to $| e \rangle$ through a virtual state  $| j_\pm \rangle$ is shown.} \label{fig:eg}
\end{figure}

We first consider an atomic system with ground state $| g \rangle$ and excited state $| e \rangle$. For the atomic systems we are interested in, for example vibrational modes of parahydrogen, and electronic states of Ytterbium, or Xenon \cite{Song:2015xaa}, both states $| g \rangle$ and $| e \rangle$ will have even parity, meaning that $E1$ dipole transitions between the two states are forbidden. However, it will be possible to excite state $| g \rangle$ to state $| e \rangle$ through multiple $E1$ dipole transitions, and similarly de-excite $| e \rangle$ to $| g \rangle$. So besides states $| g \rangle$ and $| e \rangle$, we consider intermediate states, $| j_{+}  \rangle$ and $| j_{-}  \rangle$, where $+(-)$ will indicate excitation into a positive (negative) angular momentum state by a circularly polarized photon. Figure \ref{fig:eg} illustrates the basic setup. In physical realizations, there will be many $j$ states to transition through, for example the $\ell =0,1,2,3...$ electronic angular momentum states of hydrogen. Since by design these excited states will lie at energies beyond those provided by the input lasers, transitions through these states will be virtual. Defining our atomic Hamiltonian as
\begin{eqnarray}
H = H_0 + H_I
\end{eqnarray}
 where $H_I$ is the interaction Hamiltonian and $H_0$ is defined by $H_0| g \rangle = \omega_g | g \rangle$, $H_0| e \rangle = \omega_e | e \rangle$, $H_0| j_\pm \rangle = \omega_j | j_\pm \rangle$. 
With our states specified, we define the wavefunction for this simplified atomic system
\begin{equation}
|\psi\rangle=\cg e^{-i\wg t} \g + \ce e^{-i(\we+\delta)t}\e+\cjp e^{-i\wj t}|j_+\rangle + \cjm e^{-i\wj t}|j_-\rangle\,.
\end{equation}
We have added a phase $\delta$ to account for detuning of the lasers; in other words, the laser beams exciting the atoms will be off resonance by a factor $\sim \delta$.

The laser-atomic interaction Hamiltonian will depend on the orientation and quality of the impinging laser beams. Experimental setups similar to the one we are outlining (for example \cite{Hiraki:2018jwu}), employ counter-propagating beams which have been circularly polarized. Therefore, we will consider two counter-propagating laser beams propagating along the $z$ direction with electric fields given as
\begin{align}
&\tilde{E}_{1}=\dfrac{1}{2}E_{1}(z,t)\vec{\epsilon}_l\exp\{-i\omega_1 (t+z)\}+\dfrac{1}{2}E^*_{1}(z,t)\vec{\epsilon}_r \exp\{i\omega_1 (t+z)\}\,,\label{eq:E1}\\
&\tilde{E}_{2}=\dfrac{1}{2}E_{2}(z,t)\vec{\epsilon}_l\exp\{-i\omega_2 (t-z)\}+\dfrac{1}{2}E^*_{2}(z,t)\vec{\epsilon}_r\exp\{i\omega_2 (t-z)\}\,,\label{eq:E2}
\end{align}
where $\vec{\epsilon}_{r},\vec{\epsilon}_{l}$ are unit normalized right- and left-handed polarization vectors for the laser beams. Then the laser-atom interaction Hamiltonian is
\begin{equation}
H_{I}=-\vec{d}\cdot(\tilde{E}_1+\tilde{E}_2)\,.
\end{equation}
where $\vec{d}$ is the polarization of the atom. The actual dipole coupling and transition rate are experimental inputs in these formulae. Here we define the expectation value of the dipole transitions, with the assumption that both counter-propagating pump lasers will have left-handed circular polarization, using the convention that left-handedness is defined along the direction of the beam propagation. More explicitly, since $\tilde{E}_{2}$ is electric field of a laser beam propagating in the $+\vec{z}$ direction, 
\begin{align}
d_{jg} \equiv \langle j_{+}|-\vec{d}\cdot\vec{\epsilon}_{r}|g\rangle = \langle j_{-}|-\vec{d}\cdot\vec{\epsilon}_{l}|g\rangle \nonumber \\
d_{je} \equiv \langle j_{+}|-\vec{d}\cdot\vec{\epsilon}_{r}|e\rangle = \langle j_{-}|-\vec{d}\cdot\vec{\epsilon}_{l}|e\rangle
\end{align} 
while also for $\tilde{E}_{2}$  
\begin{align}
\langle j_{+}|-\vec{d}\cdot\vec{\epsilon}_{l}|g\rangle = \langle j_{-}|-\vec{d}\cdot\vec{\epsilon}_{r}|g\rangle=0 \nonumber \\
\langle j_{+}|-\vec{d}\cdot\vec{\epsilon}_{l}|e\rangle = \langle j_{-}|-\vec{d}\cdot\vec{\epsilon}_{r}|e\rangle=0
\end{align} 
The same relations hold for $\tilde{E}_{1}$, except with $\vec{\epsilon}_{l} \leftrightarrow \vec{\epsilon}_{r} $, since $\tilde{E}_{1}$ is the electric field of the laser beam propagating in the $-\vec{z}$ direction. Finally, assuming that the two lasers will carry the same frequency (up to a detuning factor $\delta$), we define $\omega \equiv \omega_1 = \omega_2 = \omega_{eg}/2$, and we use the convention throughout this paper that $\omega_{ik}$ for any states $i,k$ is defined as $\omega_{ik} = \omega_i -\omega_k$, to arrive at the following Schr\"odinger equations for this multi-state atomic system

\begin{align}
 \label{eq:coherjp}
 i\partial_t \cjp &= \dfrac{1}{2}(\djg \cg e^{i\wjg t}+\dje \ce e^{i(\wje- \delta) t})(\Eob e^{-i\omega t}+\Etb^* e^{i\omega t})\\
  \label{eq:coherjm}
 i\partial_t \cjm &=\dfrac{1}{2}(\djg \cg e^{i\wjg t}+\dje \ce e^{i(\wje - \delta) t})(\Eob^* e^{i\omega t}+\Etb e^{-i\omega t})\\
 \label{eq:cohercg}
   i\partial_t \cg &= \dfrac{1}{2}\dgj  e^{-i\wjg t}[\cjp(\Eob^* e^{i\omega t}+\Etb e^{-i\omega t})+\cjm(\Eob e^{-i\omega t}+\Etb^* e^{i\omega t})]\\
   i\partial_t \ce &= \dfrac{1}{2}\dej  e^{-i(\wje - \delta) t}[\cjp(\Eob^* e^{i\omega t}+\Etb e^{-i\omega t})+\cjm(\Eob e^{-i\omega t}+\Etb^* e^{i\omega t})]
\label{eq:coherce}
\end{align}
where we have incorporated the spatial part of the electric fields into ``barred'' quantities, $\Eob=E_1e^{-i\omega z}$ and $\Etb=E_2e^{i\omega z}$. The sum over all intermediate states $j$ is implicit in Eqs.~\eqref{eq:coherjp}-\eqref{eq:coherce}. 

To find the time evolution of this system, we first integrate Eq.~\eqref{eq:coherjp} and Eq.~\eqref{eq:coherjm} over $t$. We will be using the so-called Markov approximation, treating $\cg,\ce$ as constant in the resulting integral. This standard approximation is justified, so long as we expect virtual transitions through $\cjp$ and $\cjm$ to be sufficiently rapid compared to changes in $\cg,\ce$, which should be satisfied so long as the frequency of the $| g \rangle \rightarrow | e \rangle$ transition is substantially smaller than the frequency of higher energy atomic states, $\omega_{eg} \ll \wje, \wjg$. 
For example, in the case of pH$_2$ the frequency of the first vibrational state, $\omega_{eg} \sim 0.5 ~{\rm eV}$, can be compared to the lowest lying electronic excitations, $\wje, \wjg \sim 10~{\rm eV}$, from which we conclude that the Markov approximation is justified. Using similar logic, we approximate the electric fields of the laser beams as being constant in this integral, since the laser frequency is also small compared to the transition frequencies to intermediate $j$ states. 
Setting the initial condition $c_{j_\pm,0}=0$, we find the time evolution of $\cjp$ and $\cjm$,
\begin{align}
\cjp &= - \frac{1}{2} \Bigg[ d_{jg} c_g \frac{e^{i (\wjg-\omega)t} -1}{\wjg-\omega} \bar{E_1} +d_{jg} c_g \frac{e^{i (\wjg+\omega)t} -1}{\wjg+\omega} \bar{E_2^*} \nonumber \\  &+d_{je} c_e \frac{e^{i (\wje-\delta-\omega)t} -1}{\wje-\delta-\omega} \bar{E_1} +d_{je} c_e \frac{e^{i (\wje-\delta +\omega)t} -1}{\wje-\delta +\omega} \bar{E_2^*}\Bigg] ,\\
\cjm &= - \frac{1}{2} \Bigg[ d_{jg} c_g \frac{e^{i (\wjg-\omega)t} -1}{\wjg-\omega} \bar{E_2} +d_{jg} c_g \frac{e^{i (\wjg+\omega)t} -1}{\wjg+\omega} \bar{E_1^*} \nonumber \\  &+d_{je} c_e \frac{e^{i (\wje-\delta-\omega)t} -1}{\wje-\delta-\omega} \bar{E_2} +d_{je} c_e \frac{e^{i (\wje-\delta +\omega)t} -1}{\wje-\delta +\omega} \bar{E_1^*}\Bigg] 
\label{eq:cjsimp}
\end{align}
Substituting these solutions for $\cjp$ and $\cjm$ into the Schrodinger equations for $\cg$ and $\ce$, we invoke the slowly varying envelope approximation, $i.e.$ we assume that since the development of the electric fields around the atoms is slow compared to the frequencies of all transitions, all time-dependent exponentials of the form 
\begin{align}
e^{i (\wje-\delta +\omega)t} \approx e^{i (\wjg+\omega)t} \approx e^{i (\wje-\delta -\omega)t} \approx e^{i (\wjg-\omega)t} \approx 0,
\end{align}  
can be set to zero. With the slowly varying envelope approximation, the two state system can be compactly expressed as
\begin{equation}
i\partial_t\left(\begin{split}
&\ce\\
&\cg\end{split}\right)=H_{eff}\left(\begin{split}
&\ce\\
&\cg\end{split}\right)\,,
\end{equation}
with the effective Hamiltonian
\begin{equation}
-H_{eff}=\left(\begin{split}
&\Wee &\Weg\\
&\Wge &\Wee\end{split}\right)\,,
\end{equation}
where $\Wge$ is the Rabi frequency of the system,
\begin{align}
\Wee=&\dfrac{\aee}{4}\left(|\Eob|^2+|\Etb|^2\right)\,,\\
\Wgg=&\dfrac{\agg}{4}\left(|\Eob|^2+|\Etb|^2\right)\,,\\
\Weg=&\Wge^*=\dfrac{\age}{2}\Eob \Etb \,,\label{eq:Weg}
\end{align}
and we have defined interstate dipole couplings as in \cite{Harris:97},
\begin{align}
&\aee=\sum\limits_j|\dje|^2\left(\dfrac{1}{\wje-\delta-\omega}+\dfrac{1}{\wje-\delta+\omega}\right)\,,\\
&\agg=\sum\limits_j|\djg|^2\left(\dfrac{1}{\wjg-\omega}+\dfrac{1}{\wjg+\omega}\right)\,,\\
&\aeg=\age^*=\sum\limits_j\dfrac{\dje\dgj}{\wje-\delta+\omega}\,.\label{eq:age}
\end{align}
Applying the density matrix of the atomic system
\begin{align}
\rho = \left( \begin{array}{cc}
| e \rangle \langle e | & | e \rangle \langle g | \\
| g \rangle \langle e | & | g \rangle \langle g | \\
\end{array} \right)= \left( \begin{array}{cc}
\rho_{ee} & \rho_{eg}  \\
\rho_{ge} & \rho_{gg}\\
\end{array} \right)
\end{align}
to the von Neumann equation $i \partial_t \rho = [H_{eff}, \rho]$, leads to the Maxwell-Bloch equations
\begin{align}
&\partial_t\ree=i(\Weg\rge-\Wge\reg)-\dfrac{\ree}{T_1}\,,\label{eq:ere}\\
&\partial_t\rge=i(\Wgg-\Wee-\delta)\rge+i\Wge(\ree-\rgg)-\dfrac{\rge}{T_2}\,,\label{eq:erg}\\
&\partial_t\rgg=i(\Wge\reg-\Weg\rge)+\dfrac{\ree}{T_1}\,.\label{eq:grg}
\end{align}
The final terms in Eqs.~\eqref{eq:ere}, \eqref{eq:erg}, and~\eqref{eq:grg} have been added to account for spontaneous $| e \rangle \rightarrow | g \rangle$ transitions and the decoherence time of the mixed state. As such, $T_1$ is the excited state lifetime and $T_2$ is the decoherence time.

To better quantify the coherence of this system, we define the Bloch vector $\vec{r}= \Tr(\vec{\sigma}\rho)$ where $\vec{\sigma}$ are the Pauli matrices. This implies
\begin{align}
&r_1=\rge+\reg\,,\\
&r_2=i(\reg-\rge)\,,\\
&r_3=\ree-\rgg\,.
\end{align}
By construction, $r_1$ and $r_2$ quantify the degree to which the atoms are coherently excited of the system, with maximum coherence attained when $r_1=1$ and $r_3=0$. The Bloch vector direction $r_3$ indicates the population difference between the excitation and the ground states. Note that $r_1$, $r_2$, and $r_3$ are all real numbers. Applying the Bloch vector basis to Eqs.~\eqref{eq:ere}, \eqref{eq:erg}, and~\eqref{eq:grg} we obtain

\begin{align}
&\partial_t r_1=\left[-\dfrac{\agg-\aee}{4}(|\bar{E}_1|^2+|\bar{E}_2|^2)+\delta\right]r_2+\aeg \im(\bar{E}_1\bar{E}_2)r_3-\dfrac{r_1}{T_2}\,,\label{eq:er1simp}\\
&\partial_t r_2=\left[\dfrac{\agg-\aee}{4}(|E_1|^2+|E_2|^2)-\delta\right]r_1+\aeg \re(\bar{E}_1\bar{E}_2)r_3-\dfrac{r_2}{T_2}\,,\label{eq:er2simp}\\
&\partial_t r_3=-\aeg[\im(\bar{E}_1\bar{E}_2)r_1+\re(\bar{E}_1\bar{E}_2)r_2]-\dfrac{1+r_3}{T_1}\,,\label{eq:er3simp}
\end{align}
where we note that $\aeg$ is assumed to be real. 

\subsection{Quantifying coherence in quasi-stable excited states}

Using the Bloch vector time evolution given by Eqs.~\eqref{eq:er1simp}-\eqref{eq:er3simp}, we can now determine the degree and duration of coherence in cold atomic preparations excited by two counter-propagating lasers with electric fields $\tilde E_1$ and $\tilde E_2$. We consider an excited set of atoms with an expected spontaneous deexcitation time (not including superradiant enhancement) of $T_1$ and a decoherence time $T_2$. In the case of the first vibrationally excited state of $\ph$, the total lifetime has been observed to be $T_1 \sim 10 ~{\rm \mu s}$ at $\sim 10~{\rm K}$ temperatures \cite{Kuo:1984,li2001measurement,DELALANDE1977339}, which will be appreciably longer than the decoherence time of the first $\ph$ vibrational excitation at these temperatures, where this decoherence time will be of order $\sim 1-100~{\rm ns}$ \cite{Bischel:1986}. 

In more detail, the decoherence time ($T_2$) of $\ph$ has been studied extensively for a variety of temperatures and densities \cite{Bischel:1986,Fukumi:2012rn,2017JPCA..121.3943M,Hiraki:2018jwu}. In some regimes, it is accurate to use the mean interaction time of hydrogen atoms as an estimate of the decoherence time resulting from pH$_2$ collisions at number density $n$,
\begin{align}
t_{dc} &= \frac{1}{n_{H} \sigma_{H} \sqrt{2T/m_{H}}} \nonumber \\&\approx 3 ~{\rm ns} \left( \frac{3 \times 10^{19} ~{\rm cm^{-3}}}{n} \right)  \left( \frac{9 \times 10^{-17} ~{\rm cm^2}}{\sigma_H} \right)
  \left( \frac{80 ~{\rm K}}{T} \right)^{1/2}  \left( \frac{m_H}{0.94~{\rm GeV}} \right)^{1/2},
  \label{eq:gascoh}
\end{align}
where this expression has approximated the velocity of pH$_{2}$ using the temperature ($T$) and mass ($m_H$) of hydrogen, and the cross-section using the Bohr radius, $\sigma_H \approx \pi r_{bohr}^2$. 

\begin{table}[h!]
\centering
 \begin{tabular}{||c c c c||} 
 \hline
 $\ph$ Reference &  Density (cm$^{-3}$) & Temperature (K) & Decoherence Time (ns)  \\ [0.5ex] 
 \hline\hline
  \cite{Bischel:1986} &$ 10^{19}-10^{20}$ & 80-500 & $\sim 10$ \\ [.1ex] \hline
\cite{Miyamoto:2014} & $5.6 \times 10^{19}$ & 78 & $\sim 8$  (est) \\  [.1ex]\hline
\cite{Hiraki:2018jwu} &  $10^{19} -  5 \times 10^{20}$ & 78 & $\sim 10$ (est)  \\ [.1ex] \hline
 \cite{momose2005high} & $2.6 \times 10^{22}$ & 4.2 & $\gtrsim 140$ \\ [.1ex]
 \hline
 \end{tabular}
 \caption{A number of parahydrogen experiments with long decoherence times are listed for comparison, along with their temperatures and number densities. For $\ph$ Raman linewidth measurements $\Delta \nu_{dec}$, the decoherence times are estimated as $T_2 \sim \frac{1}{\Delta \nu_{dec}}$. In the case of References \cite{Miyamoto:2014,Hiraki:2018jwu}, the decoherence time is an estimate using results from Ref.~\cite{Bischel:1986}.}
 \label{tab:decoh}
\end{table}
While Eq.~\eqref{eq:gascoh} is remarkably close to the measured decoherence time for a sample of $\ph$ prepared at $T \sim 80$ Kelvin and density $n \sim 3 \times 10^{19}~{\rm cm^{-3}}$, this approximation will break down for sufficiently cold and dense pH$_2$, which will not behave like an ideal gas. In addition, we should note that the Raman linewidth, or full-width-half-at-half-maximum of $\ph$'s first vibrational emission line, is often used to determine the decoherence time. However, this linewidth also has a contribution from Doppler broadening of $\ph$
\begin{equation}
\Delta \nu_{dec}^{(Doppler)} \approx \omega_0 \sqrt{\frac{T}{m_H}}, 
\end{equation} 
where here $\omega_0$ is the first vibrational mode frequency.
A total decoherence determination for the first vibrational mode of $\ph$, for temperatures ranging from $77-500$ K, was approximated by fitting a phenomenological formula \cite{Bischel:1986}
\begin{align}
\Delta \nu_{dec} = \frac{A} {n} + B n,
\end{align}
where, for example, it was found that for $T= 80$ K, the collisional term $A \approx 100 ~{\rm MHz~cm^{-3}}$, and the broadening term $B \approx 20 ~{\rm MHz~cm^{3}}$, which implies a 10 ns decoherence time for $n \sim 10^{19}~{\rm cm^{-3}}$, as previously noted.

\begin{figure}[!htb]
\centering
\includegraphics[width=1.\textwidth]{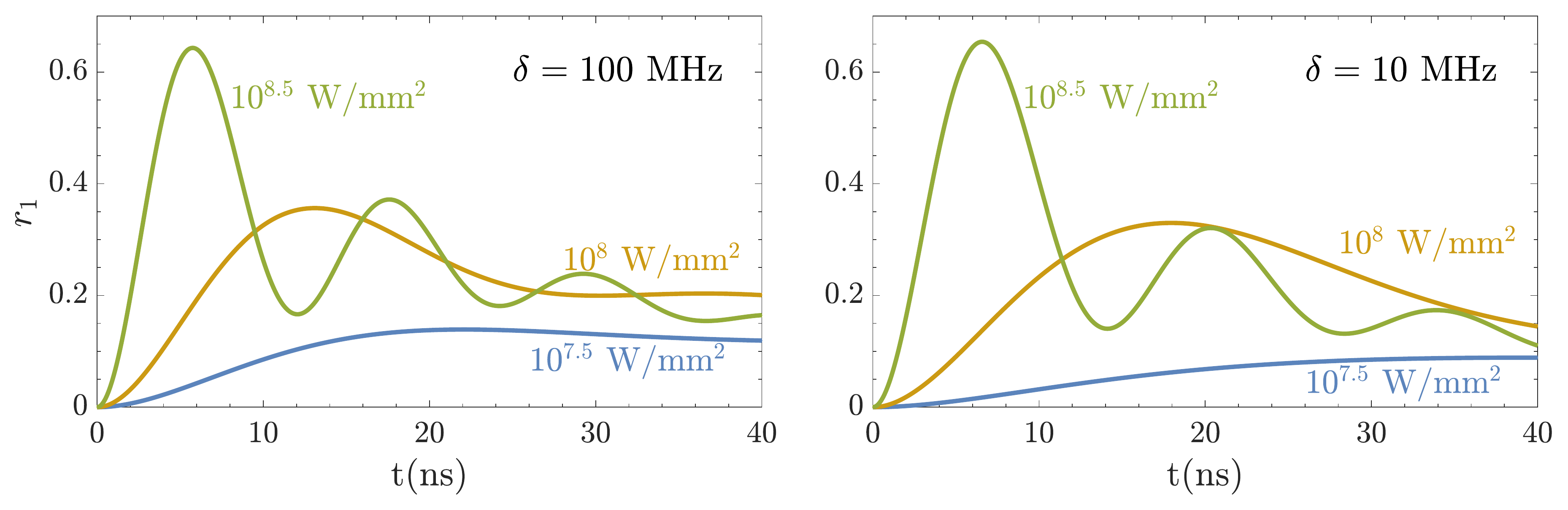}
\caption{The development of coherence in parahydrogen pumped by two counter-propagating lasers tuned to half the frequency of the first vibrational state of parahydrogen, $\omega = 0.26 $ eV, with an assumed parahydrogen number density of $n = 10^{21} {\rm ~cm^{-3}} $. Results were obtained by solving Eqs.~\eqref{eq:er1simp}-\eqref{eq:er3simp}, where we take $\agg=0.90 \times 10^{-24}\ \mathrm{cm}^3$, $\aee=0.87 \times 10^{-24}\ \mathrm{cm}^3$, $\aeg=0.0275\times 10^{-24}\ \mathrm{cm}^3$ \cite{Fukumi:2012rn}. The intensity of the pump lasers is indicated. For comparison, we note that a coherence of $r_1 \simeq 0.07$ has been achieved for parahydrogen at density $5 \times 10^{19} {\rm ~cm^{-3}} $, using lasers less powerful than those assumed here \cite{Miyamoto:2015tva}. However, the nanosecond pulse gigawatt power lasers required are commerically available \cite{laser}. (Indeed, even continuous gigawatt lasers as powerful as we require have been demonstrated in recent years \cite{2016arXiv161008493S}.) The left panel assumes experimental detuning $\delta = 100$ MHz, as achieved in recent counter-propagating pulsed laser experiments \cite{Hiraki:2018jwu}. The right panel assumes $\delta =10$ MHz, a linewidth that has been achieved in solid parahydrogen \cite{momose2005high}.} \label{fig:decoh}
\end{figure}

Given the theoretical expectations and experimental results detailed above, it is safe to assume that $T_2 \approx 10~{\rm ns}$ is an achievable decoherence time for cold parahydrogen. In terms of Bloch vector $r_1$, the largest coherence reported in a similar setup was $r_1 \simeq 0.07$ for parahydrogen at density $n \sim 5 \times 10^{19} {\rm ~cm^{-3}} $ \cite{Miyamoto:2015tva}.  In the remainder of this article, we will find that advancing coupling sensitivity to dark photons (assuming a roughly 30 cm cylindrical chamber and 1 cm laser beam diameter) requires parahydrogen number densities nearer to $n \sim 10^{21} {\rm ~cm^{-3}} $. As noted in Figure \ref{fig:decoh}, a higher-power laser than that used in \cite{Miyamoto:2015tva} is also required. In Figure \ref{fig:decoh}, we have shown how coherence of pH$_2$ can be expected to develop in time for $n \sim 10^{21} {\rm ~cm^{-3}} $, by solving Eqs.~\eqref{eq:er1simp}-\eqref{eq:er3simp}, assuming a $\sim 10$ nanosecond decoherence time, and intrinsic detuning by experimental effects like Doppler broadening, of both $\delta = 10$ and $\delta = 100~{\rm MHz} $. We will see that in this $\sim$ ten nanosecond timeframe, a dark photon field applied to the cold atoms can greatly enhance the two photon transition rate.

\section{Dark photons in two stage atomic transitions}
\label{sec:darkSR}
We have found that substantial coherence can be established in atoms excited by counter-propagating lasers, through a two photon excitation process. Similarly, in the presence of a dark photon field, the rate for two photon de-excitation can be resonantly enhanced. Suitably applied to coherently excited atoms, we will find that very weakly coupled dark photon fields can trigger two photon transitions, during the $\sim 10$ nanosecond window of time that the atoms are coherently excited.

\subsection{Two photon transitions with kinetic mixing}
We begin with the dark photon. The dark photon field is a new massive $U(1)$ gauge field that kinetically mixes with the Standard Model photon. Its Lagrangian has the general form
\begin{equation}
\mathcal{L}=-\frac{1}{4}(F_{\mu\nu}F^{\mu\nu}-2\chi F_{\mu\nu}F'^{\mu\nu}+F'_{\mu\nu}F'^{\mu\nu})+\frac{\mAp^2}{2}A'_\mu A'^\mu-eJ^\mu_{\textrm{em}}A_\mu\,,
\label{eq:DPlag}
\end{equation}
where $A^\mu$ and $A'^\mu$ are the four vector potential of the ordinary photon and dark photon field, and $F_{\mu\nu}$ and $F'^{\mu\nu}$ describe their field strength separately. Additionally, the dark photon is characterized by a mass $\mAp$ and the kinetic mixing is suppressed by a constant $\chi$. Here $J^\mu_{\textrm{em}} = \bar \psi \gamma^\mu \psi $ corresponds to the electromagnetic charged current with charged fermions $\psi$.

There is no direct coupling between the dark photon and charged fermions in Eq.~\eqref{eq:DPlag}. Rather, an effective interaction is introduced through kinetic mixing between the photon and dark photon, so long as $\mAp > 0$. Equivalently, one can diagonalize the kinetic mixing term by redefinition of the photon field $A_\mu\,\rightarrow\, A_\mu+\chi A'_\mu $. To first order in $\chi$ we obtain the Lagrangian
\begin{equation}
\mathcal{L}=-\frac{1}{4}(F_{\mu\nu}F^{\mu\nu}+F'_{\mu\nu}F'^{\mu\nu})+\frac{\mAp^2}{2}A'_\mu A'^\mu-e(A_\mu+\chi A'_\mu) J^\mu_{\textrm{em}}\,,
\label{eq:DPlag2}
\end{equation}
To find the dark photon absorption and emission amplitude in atomic transitions, it will be convenient to work with the effective Hamiltonian for electrons in the non-relativistic limit in the presence of the dark photon field. Substituting the interaction terms in Eq.~\eqref{eq:DPlag2} into the Dirac Lagrangian,
\begin{equation}
\mathcal{L}= i \bar \psi \gamma^\mu \partial_\mu \psi -m_e \bar \psi \psi -e(A_\mu+\chi A'_\mu) J^\mu_{\textrm{em}}
\label{eq:DiracL}
\end{equation} 
we arrive at the Dirac equation for the electron
\begin{equation}
[i\slashed{\partial}-e(\slashed{A}+\chi \slashed{A}')-m_e]\psi=0\,,
\label{eq:Dirac}
\end{equation} 
where $m_e$ is the electron mass. 

It will be convenient to work in the Dirac basis and divide the spinor into a dominant component $\psi_d$ and a subdominant component $\psi_s$, i.e. $\psi=(\psi_d,\psi_s)^T$. Separating out the time derivative from the Dirac equation, we find the Hamiltonian for the system
\begin{equation}
i\partial_t\left(\begin{split}
&\psi_d\\
&\psi_s
\end{split}\right)=H\left(\begin{split}
&\psi_d\\
&\psi_s
\end{split}\right)\,,
\label{eq:Schrodinger}
\end{equation}
where
\begin{equation}
H=\left(\begin{split}
&e(\Phi+\chi A'_0)+m_e & -i\vec{\sigma}\cdot\vec{\nabla}-e\vec{\sigma}\cdot(\vec{A}+\chi \vec{A'})\\
&-i\vec{\sigma}\cdot\vec{\nabla}-e\vec{\sigma}\cdot(\vec{A}+\chi \vec{A'}) &e(\Phi+\chi A'_0)-m_e 
\end{split}\right)\,,
\end{equation}
Here $\vec \sigma$ are the Pauli spin matrices and the electric potential $\Phi=A_0$. 

The non-relativistic Hamiltonian for this system is obtained by subtracting $m_e$ from both sides of Eq.~\eqref{eq:Schrodinger}, which yields
\begin{align}
H_{nr}\psi_d&=-[i\vec{\sigma}\cdot\vec{\nabla}+e\vec{\sigma}\cdot(\vec{A}+\chi \vec{A'})]\psi_s+e(\Phi+\chi A'_0)\psi_d\label{eq:psid}\\
H_{nr}\psi_s&=-[i\vec{\sigma}\cdot\vec{\nabla}+e\vec{\sigma}\cdot(\vec{A}+\chi \vec{A'})]\psi_d+e(\Phi+\chi A'_0)\psi_s-2m_e\psi_s\label{eq:psis}\,.
\end{align}
The subdominant component $\psi_s$ can be solved in the non-relativistic limit where $|H_{nr}| \ll m_e$ and $|e(\Phi + A'_0)| \ll m_e$. It is then substituted into Eq.~\eqref{eq:psid} to obtain
\begin{equation}
H_{nr}=H-m_e=\frac{1}{2m_e}[i\vec{\sigma}\cdot\vec{\nabla}+e\vec{\sigma}\cdot(\vec{A}+\chi \vec{A'})]^2+e(\Phi+\chi A'_0)\,,
\label{eq:Hamiltonianall}
\end{equation}
where this expression is valid in the non-relativistic limit where 
Eq.~\eqref{eq:Hamiltonianall} gives the effective Hamiltonian for an electron in the presence of electromagnetic and dark photon fields. Subtracting from it the standard QED Hamiltonian we single out the components introduced by the new dark photon field
\begin{equation}
\begin{split}
H_{A'}=&\frac{e\chi}{2m_e}[i(\vec{\nabla}\cdot \vec{A'}+\vec{A'}\cdot\vec{\nabla})]-\frac{e\chi}{2m_e}\vec{\sigma}\cdot(\vec{\nabla}\times\vec{A'})+e\chi A'_0\\
&+\frac{e^2\chi}{m_e}\vec{A}\cdot\vec{A'}+\frac{e^2\chi^2}{2m_e}\vec{A'}^2\,.
\end{split}
\label{eq:HamiltonianDP}
\end{equation}
The first line of Eq.~\eqref{eq:HamiltonianDP} reminds us of the standard QED Hamiltonian, with an additional gauge field. The $\frac{e^2\chi}{m_e}\vec{A}\cdot\vec{A'}$ terms arises from the expansion of the bracket in Eq.~\eqref{eq:Hamiltonianall}, meaning that even if the kinetic mixing is explicitly removed by a specific gauge choice in Eq.~\eqref{eq:DPlag2}, dark photon and photon fields can still act on electron in a collective way. 

With the effective Hamiltonian in hand we are now prepared to compute the transition amplitude from initial atomic state $|i\rangle$ to final state $|f\rangle$ with the absorption or emission of a dark photon. This transition has the general form
\begin{equation}
\calM=\langle f|H_{A'}|i\rangle\,.
\end{equation}

We start with the first term in Eq.~\eqref{eq:HamiltonianDP}, which describes an $E1$ (electric-dipole) type transition. Using the relation $\partial_\mu A'^\mu=0$, which can be readily obtained from the Euler-Lagrange equation \eqref{eq:DiracL} for the dark photon, we find
\begin{equation}
H_{A'}^{E1}=i\frac{e\chi}{2m_e}(\vec{\nabla}\cdot \vec{A'}+\vec{A'}\cdot\vec{\nabla})=-\frac{e\chi}{2m_e}(i\partial_t A'_0+2\vec{A'}\cdot \vec{p}_e)\,,
\end{equation}
where $\vec{p}_e$ is the momentum operator for the electron. Using the relation $\vec{p}_e=-i m_e[\vec{r},\,H_0]$ where $H_0 = p_e^2/2m_e$ is the unperturbed atomic Hamiltonian, we obtain
\begin{equation}
\calM^{E1}=-\frac{e\chi}{2m_e}\langle f|\partial_t A'_0|i\rangle+i\omega_{if}e\chi \langle f|\vec{A'}\cdot \vec{r}|i\rangle\,,
\label{eq:ME1}
\end{equation}
where again we note that $\omega_{ik} \equiv \omega_i-\omega_k$ as the energy difference between the initial and final atomic states. The first term in Eq.~\eqref{eq:ME1} is suppressed by a factor $\sim \omega/m_e$ and is therefore negligible compared to the second term. Hence we drop this first term for simplicity. 

To evaluate the second term, we define the vector component of the dark photon field as $\vec{A'}=|\vec A'|\vec{\epsilon}'\exp( i\omega t- i\vec{k}\cdot \vec{r})$, which will have energy $\omega=\omega_{if}$. Because we will be considering dipole moments substantially smaller than the wavelength of the applied laser (or the wavelength of the dark photon), the dipole approximation $\exp(- i\vec{k} \cdot \vec{r})\simeq 1$ applies. With this approximation
\begin{equation}
\calM^{E1}\simeq i e\chi\omega_{if} |\vec A'|\langle f|\vec{\epsilon}\cdot \vec{d}|i\rangle\,,
\label{eq:ME1pol}
\end{equation}
where the $\vec{d}=e\vec{r}$ is the dipole operator. Following standard electromagnetic conventions we define the dark electric field as
\begin{equation}
\vec{E}'=-\vec{\nabla}V'-\partial_t \vec{A}'\,,
\label{eq:darkE}
\end{equation}
where
\begin{equation}
V'(\vec{r},t)=\dfrac{i}{\omega}\vec{\nabla}\cdot\vec{A}'\,.
\end{equation}
Assuming $|A'|$ varies slowly in space and time, we obtain
\begin{equation}
\vec{E}'=\dfrac{i}{\omega}[(\vec{k}\cdot\vec{A}')\, \vec{k}-\omega^2\vec{A}']\,.
\end{equation}
Decomposing the dark electric field into a transverse component $\vec{E}'_T$ and longitudinal component $\vec{E}'_L$,
\begin{align}
&\vec{E}'_T=-i\omega \vec{A'}_T\,,\\
&\vec{E}'_L=-i\dfrac{\mAp^2}{\omega}\vec{A'}_L\,.
\end{align}
If $|\vec{A'}_T|\simeq |\vec{A'}_L|$, we have $|\vec{E'}_L|/|\vec{E'}_T|\simeq \mAp^2/\omega^2$. Note that our proposed experiment is only sensitive to dark photons with sub-meV masses, since this is a necessary condition for coherent excitation of two stage atomic transitions in the target sample (see Section \ref{sec:sens}). While the dark photon masses will be $m_{A'} \lesssim {\rm meV}$, the transition energy $\omega\sim eV$, therefore we expect the effect of longitudinal component of the dark electric field to be subdominant since $|\vec{E'}_L|/|\vec{E'}_T|\simeq \mAp^2/\omega^2$.  Therefore, we only focus on the transverse component in computing the transition amplitude,
\begin{equation}
\calM^{E1}\simeq -\chi \langle f|\vec{d}\cdot \vec{E'_T}|i\rangle\,.
\end{equation}
We could also evaluate the transition amplitude induced by other terms in Eq.~\eqref{eq:HamiltonianDP}. However, we note that the second term is characterized by M1 (magnetic dipole) type transition which is suppressed by $1/m_e$ compared with $\calM^{E1}$. The third term vanishes in the leading order expansion of $\exp(-i\vec{k}\cdot \vec{r})$. The fourth and last terms are also suppressed by $1/m_e$, and the last term is further suppressed by $\chi$, and so we also drop these terms.

\subsection{Dark photon induced superradiance}

\begin{figure}[!htb]
\centering
\begin{subfigure}{0.31\textwidth}
\includegraphics[width=\linewidth]{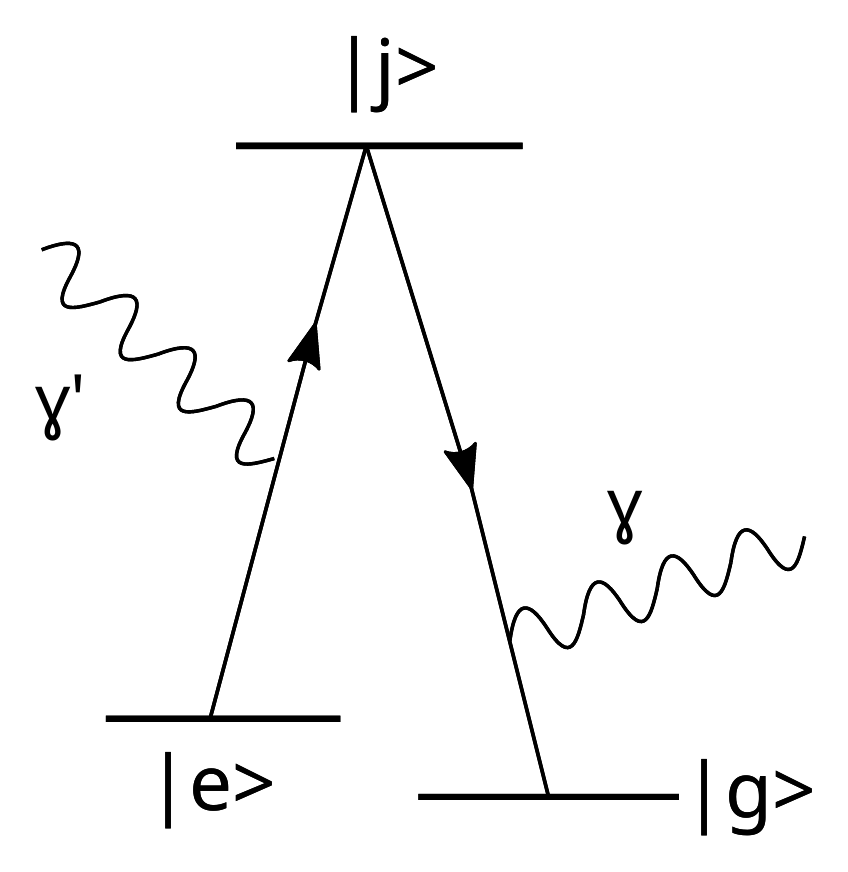}
\caption{$|e\rangle\rightarrow |g\rangle+\gamma'+\gamma$} \label{fig:transit_a}
\end{subfigure}
\hspace*{0.25cm} 
\begin{subfigure}{0.31\textwidth}
\includegraphics[width=\linewidth]{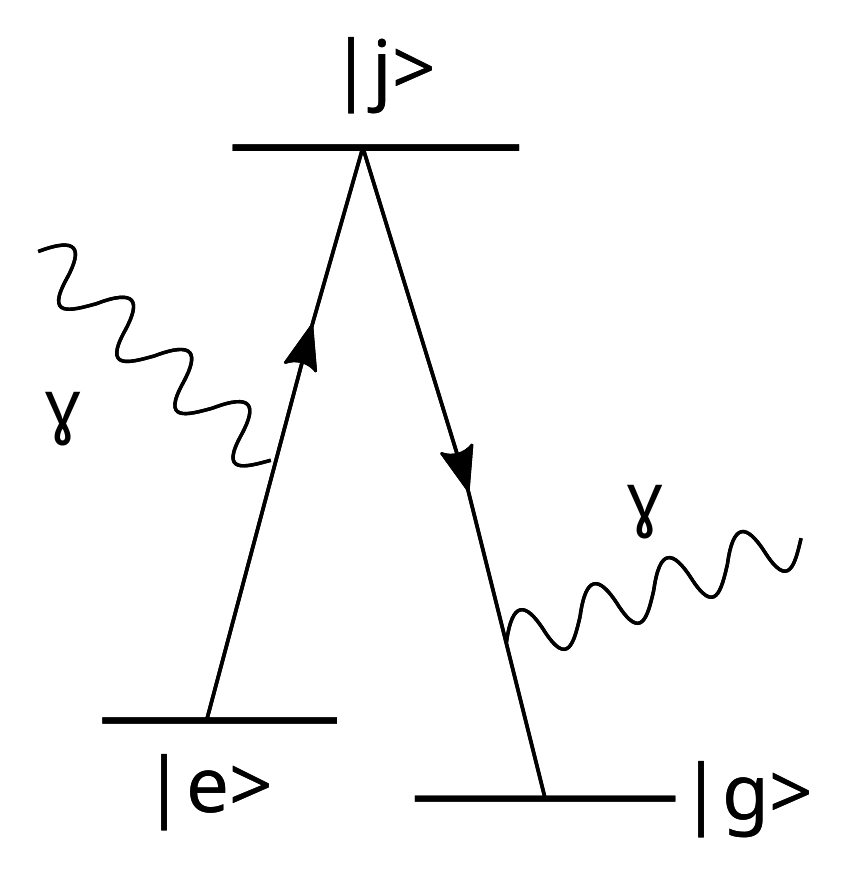}
\caption{$|e\rangle\rightarrow |g\rangle+\gamma+\gamma$} \label{fig:transit_b}
\end{subfigure}
\caption{Illustration of the two deexcitation processes. Left panel: The transition from $|e\rangle$ to $|g\rangle$ with the emission of a photon and a dark photon. Right panel: The transition from $|e\rangle$ to $|g\rangle$ with the emission of two photons.}
\end{figure}

Now that we have obtained the dark photon dipole transition amplitude, we are ready to study dark photon induced superradiance. We will focus on the transition between the excitation state $|e\rangle$ and ground state $|g\rangle$ of a $\ph$ target. As previously noted, $|g\rangle$ and $|e\rangle$ will denote the 0th and 1st vibrational state for $\ph$ where both of these have $J=0$. Since $\e$ and $\g$ share the same parity, an $E1$ dipole transition is forbidden, but the transition between them can take place via two $E1$ transitions, by transitioning through an intermediate virtual state $|j\rangle$. Hence we will compute $E1\times E1$ transitions for which two particles are emitted, as shown in Figures~\ref{fig:transit_a} and \ref{fig:transit_b}. As we mentioned in the previous section the interaction between dark photon and electron allows for this  $E1\times E1$ transition to occur via the emission of a dark photon and standard model photon $|e\rangle\rightarrow |g\rangle+\gamma'+\gamma$ along with the standard two photon emission process $|e\rangle\rightarrow |g\rangle+\gamma+\gamma$, illustrated in Figures~\ref{fig:transit_a} and \ref{fig:transit_b}. These two processes will reinforce each other in a coherently excited atomic medium, since the emission of a dark photon can trigger and amplify the two photon emission process, and vice versa. To demonstrate this mutual reinforcement, we shall derive the evolution equations of the dark photon and photon fields during deexcitation.

\subsubsection{Maxwell-Bloch equations}
\label{sec:maxwelleq}
First we will reformulate the Maxwell-Bloch equations as they were derived out in Section \ref{sec:2state}, now including the dark photon's effect on the electric dippole. As before, we denote the spin $m_J=\pm 1$ states as $|j_{\pm}\rangle$. In addition to the two photon fields $E_{1}$ and $E_{2}$, we define a dark photon $E'$ propagating in the positive $z$ direction 
\begin{align}
&\tilde{E}'=\dfrac{1}{2}E'(z,t)\vec{\epsilon}'\exp\{-i\omega' t+ikz\}+c.c.\,.
\end{align}
Because we are only treating the transverse component of the dark photon field, we take $\vec{\epsilon}'=\vec{\epsilon}'_T$. We expect that to good approximation $\omega_1=\omega_2=\omega'=\omega =\weg/2$, since this is already required for coherence of the excited atomic state. We again write the $\ph$ wave function as the superposition of atomic states
\begin{equation}
|\psi\rangle=\cg e^{-i\wg t} \g + \ce e^{-i(\we+\delta)t}\e+\cjp e^{-i\wj t}|j_+\rangle + \cjm e^{-i\wj t}|j_-\rangle\,.
\end{equation}
For the sake of simplicity, we will keep $\delta$ in the derivation, but set $\delta = 0$ in the numerical simulations, which amounts to assuming that the atoms, lasers, and dark photon field are in phase over the target volume for timescales shorter than the decoherence time, $T_2 \sim 10~{\rm ns}$. For a full discussion of the physical requirements for $\sim 10$ ns decoherence times, and the loss of coherence as $\delta$ is increased, see Section \ref{sec:2state}. For discussions of the detuning effect on the output photon flux and the experimental sensitivity, see Appendix~\ref{app:detuning}. The full interaction Hamiltonian is then
\begin{equation}
H_{I}=-\vec{d}\cdot(\tilde{E}_1+\tilde{E}_2+\chi\tilde{E}')\,.
\end{equation}
The Schr\"odinger equations will now include terms proportional to the dark photon field,
\begin{align}
 i\partial_t \cjp &= \dfrac{1}{2}(\djg \cg e^{i\wjg t}+\dje \ce e^{i(\wje-\delta) t})(\Eob e^{-i\omega t}+\Etb^* e^{i\omega t})\nonumber\\
&+\dfrac{\chi}{2}(\djgp \cg e^{i\wjg t}+\djep \ce e^{i(\wje-\delta) t})(\Epb e^{-i\omega t}+\Epb^* e^{i\omega t})\,,\label{eq:dcj+}\\ 
 i\partial_t \cjm &=\dfrac{1}{2}(\djg \cg e^{i\wjg t}+\dje \ce e^{i(\wje-\delta) t})(\Eob^* e^{i\omega t}+\Etb e^{-i\omega t})\nonumber\\
 &+\dfrac{\chi}{2}(\djgp \cg e^{i\wjg t}+\djep \ce e^{i(\wje-\delta) t})(\Epb^* e^{i\omega t}+\Epb e^{-i\omega t})\,,\label{eq:dcj-}\\
   i\partial_t \cg &= \dfrac{1}{2}\dgj  e^{-i\wjg t}[\cjp(\Eob^* e^{i\omega t}+\Etb e^{-i\omega t})+\cjm(\Eob e^{-i\omega t}+\Etb^* e^{i\omega t})]\nonumber\\
   &+\dfrac{\chi}{2}\dgjp  e^{-i\wjg t}[\cjp(\Epb^* e^{i\omega t}+\Epb e^{-i\omega t})+\cjm(\Epb e^{-i\omega t}+\Epb^* e^{i\omega t})]\,,\label{eq:dcg}\\
   i\partial_t \ce &= \dfrac{1}{2}\dej  e^{-i(\wje-\delta) t}[\cjp(\Eob^* e^{i\omega t}+\Etb e^{-i\omega t})+\cjm(\Eob e^{-i\omega t}+\Etb^* e^{i\omega t})]\nonumber\\
   &+\dfrac{\chi}{2}\dejp  e^{-i(\wje-\delta) t}[\cjp(\Epb^* e^{i\omega t}+\Epb e^{-i\omega t})+\cjm(\Epb e^{-i\omega t}+\Epb^* e^{i\omega t})]\,,\label{eq:dce}
\end{align}
where as in Section \ref{sec:2state} we absorb spatial dependence into overbarred fields, $\Eob=E_1e^{-i\omega z}$, $\Etb=E_2e^{i\omega z}$ and $\Epb=E'e^{ik z}$. Note also that we have left implicit the sum over all intermediate states $j$ in Eq.~\eqref{eq:dcg} and Eq.~\eqref{eq:dce}. Integrating Eq.~\eqref{eq:dcj+} and Eq.~\eqref{eq:dcj-} over $t$, using the Markovian approximation, and imposing the initial condition $c_{j_\pm,0}=0$,
\begin{align}
\cjp=-\dfrac{1}{2}\sum\limits_{s=g,e} & \left[\dfrac{c_s}{\omega_{js}-\omega-\Delta_{se}\delta}(d_{js}\Eob+\chi d'_{js}\Epb)\left(e^{i(\omega_{js}-\omega-\Delta_{se}\delta)t}-1\right)\right.\nonumber\\
&\left.+\dfrac{c_s}{\omega_{js}+\omega-\Delta_{se}\delta}(d_{js}\Etb^*+\chi d'_{js}\Epb^*)\left(e^{i(\omega_{js}+\omega-\Delta_{se}\delta)t}-1\right)\right]\,,\label{eq:cj+}\\
\cjm=-\dfrac{1}{2}\sum\limits_{s=g,e} & \left[\dfrac{c_s}{\omega_{js}-\omega-\Delta_{se}\delta}(d_{js}\Etb+\chi d'_{js}\Epb)\left(e^{i(\omega_{js}-\omega-\Delta_{se}\delta)t}-1\right)\right.\nonumber\\
&\left.+\dfrac{c_s}{\omega_{js}+\omega-\Delta_{se}\delta}(d_{js}\Eob^*+\chi d'_{js}\Epb^*)\left(e^{i(\omega_{js}+\omega)t-\Delta_{se}\delta}-1\right)\right]\,,\label{eq:cj-}
\end{align}
where $\Delta_{se}=0$ if $s=g$ and $\Delta_{se}=1$ if $s=e$. Applying Eq.~\eqref{eq:cj+} and Eq.~\eqref{eq:cj-} in Eq.~\eqref{eq:dcg} and Eq.~\eqref{eq:dce} and using the slowly varying envelope approximation, we obtain the equation for the two-state system in the presence of a dark photon,
\begin{equation}
i\partial_t\left(\begin{split}
&\ce\\
&\cg\end{split}\right)=-\left(\begin{split}
&\Wee &\Weg\\
&\Wge &\Wee\end{split}\right)\left(\begin{split}
&\ce\\
&\cg\end{split}\right)\,,
\end{equation}
where the $2 \times 2$ matrix is the effective Hamiltonian ($H_{eff}$), and its components are
\begin{align}
\Wee=&\dfrac{\aee}{4}(|\Eob+\chi\eta\Epb|^2+|\Etb+\chi\eta\Epb|^2)\,,\\
\Wgg=&\dfrac{\agg}{4}(|\Eob+\chi\eta\Epb|^2+|\Etb+\chi\eta\Epb|^2)\,,\\
\Weg=&\Wge^*=\dfrac{\age}{2}(\Eob+\chi\eta\Epb)(\Etb+\chi\eta\Epb)\,,\label{eq:Wegdp}
\end{align}
where we have defined the dipole couplings $\aee,~\agg,$ and $\age$ as before. In contrast to Section \ref{sec:2state}, we now also define
\begin{equation}
\eta\equiv\dfrac{\djgp}{\djg}=\dfrac{\djep}{\dje}\label{eq:eta}
\end{equation}
which quantifies the relative phase between the polarization of the photon field and the dark photon field.

As before we introduce the density matrix and add relaxation terms to obtain the Maxwell-Bloch equations
\begin{align}
&\partial_t\ree=i(\Weg\rge-\Wge\reg)-\dfrac{\ree}{T_1}\,,\label{eq:eree}\\
&\partial_t\rge=i(\Wgg-\Wee-\delta)\rge+i\Wge(\ree-\rgg)-\dfrac{\rge}{T_2}\,,\label{eq:erge}
\end{align}
where $T_1$ and $T_2$ are relaxation and decoherence time, respectively. 

We can expand Eq.~\eqref{eq:Wegdp} to make manifest oscillations in $\Weg$
\begin{equation}
\Weg=\dfrac{\aeg}{2}\left[(E_1E_2+\chi\eta E_1E')+\chi\eta (E'E_2+\chi\eta E'^2)e^{2i\omega z}\right]\,,\label{eq:wegdcomp}
\end{equation}
here assuming $\omega\simeq k$. From Eq.~\eqref{eq:erge} we also need to decompose $\rge$ correspondingly
\begin{equation}
\rge=\rge^0+\rge^-e^{-2i\omega z}\,.\label{eq:rgedcomp}
\end{equation}
We note that $\rge^-$ only comes from the atomic excitation due to the absorption of $E_2$ and $E'$ or two dark photons, and the coherence developed in these processes is small. Thus, to leading order we can drop the second term in Eq.~\eqref{eq:rgedcomp} and assume no spatial phase in $\rge$.

\subsubsection{Field equations}
The Bloch equations we have derived in the previous section show the evolution of the population of ground state and the excitation state in the presence of electric and dark electric fields. Now we would like to see how these fields evolve as the population changes. It is straightforward to obtain from Eq.~\eqref{eq:DPlag2} the field equations
\begin{align}
&(\partial_t^2-\partial_z^2)A^\mu=eJ^\mu_{\textrm{em}}\,,\label{eq:fieldA}\\
&(\partial_t^2-\partial_z^2+\mAp^2)A'^\mu=e\chi J^\mu_{\textrm{em}}\,.\label{eq:fieldA'}
\end{align}
There is no free electric charge in the target and $\vec{J}_{\textrm{em}}$ can be identified as the polarization current density determined by the polarization field
\begin{equation}
e\vec{J}_{\textrm{em}}=n\dfrac{\partial\tP}{\partial t}\,,
\end{equation}
where $n$ is the number density of $\ph$. We recall the definition of $E'$ in Eq.~\eqref{eq:darkE} and take the time derivative on both sides of Eq.~\eqref{eq:fieldA} and Eq.~\eqref{eq:fieldA'} to obtain 
\begin{align}
&(\partial_t^2-\partial_z^2)\tilde{E}_i=-n\partial_t^2\tP_i\,,\label{eq:eEi}\\
&(\partial_t^2-\partial_z^2+\mAp^2)\tilde{E'}=-\chi n\partial_t^2\tP'\,,\label{eq:eE'}
\end{align}
where $i=1,2$ represent different electric fields. The polarization field arises from the dipole moment in the atomic transition where
\begin{equation}
\tilde{P}=\langle \psi|\vec{d}|\psi\rangle\,.
\end{equation}
Note that $\tilde{E}_1$ and $\tilde{E}_2$ propagate in opposite directions with opposite spin angular momenta, the microscopic polarization that sources these fields is also different. Accounting for the conservation of angular momentum we have
\begin{align}
&-\tP_1=\sum\limits_{s=g,e}(d_{sj}c_s^*\cjp e^{-i(\omega_{js}-\omega-\Delta_{se}\delta)t}+d_{js}\cjm^*c_s e^{i(\omega_{js}-\omega-\Delta_{se}\delta)t})\vec{\epsilon}_l+c.c.\,,\label{eq:eP1}\\
&-\tP_2=\sum\limits_{s=g,e}(d_{sj}c_s^*\cjp e^{-i(\omega_{js}-\omega-\Delta_{se}\delta)t}+d_{js}\cjm^*c_s e^{i(\omega_{js}-\omega-\Delta_{se}\delta)t})\vec{\epsilon}_r+c.c.\,,\label{eq:eP2}\\
&-\tP'=\sum\limits_{s=g,e}[d'_{sj} c_s^*(\cjp+\cjm)e^{-i(\omega_{js}-\omega-\Delta_{se}\delta)t}+d'_{js} c_s(\cjp^*+\cjm^*)e^{i(\omega_{js}-\omega-\Delta_{se}\delta)t}]\vec{\epsilon}'+c.c.\,.\label{eq:eP'}
\end{align}
We work in the limit where $\mAp \ll \omega$ so approximately $\omega\simeq k$. We can substitute $c_{j\pm}$ as given in Eq.~\eqref{eq:cj+} and Eq.~\eqref{eq:cj-} into \cref{eq:eP1,eq:eP2,eq:eP'} and keep only the terms containing $e^{\pm i\omega t}$ (to match the left hand side of Eq.~\eqref{eq:eEi} and Eq.~\eqref{eq:eE'}) to obtain
\begin{align}
2\tP_{1(2)}=&\left\{[(\aee\ree+\agg\rgg)E_1+2\aeg\rge^*(E_2^*+\chi\eta E'^*)]e^{-i\omega(t+z)}\right.\nonumber\\
&+[(\aee\ree+\agg\rgg)(E_2^*+\chi\eta E'^*)+2\aeg^*\rge E_1]e^{i\omega(t-z)}\nonumber\\
&+\left. [(\aee\ree+\agg\rgg)e^{-i\omega(t-z)}+2\aeg^*\rge e^{i\omega(t+z)}]\chi\eta E'\right\}\vec{\epsilon}_{l(r)}\nonumber\\
&+c.c.\,,\\
2\tP'=&\left\{[(\aee\ree+\agg\rgg)E_1+2\aeg\rge^*(E_2^*+2\chi\eta E'^*)]e^{-i\omega(t+z)}\right.\nonumber\\
&+[(\aee\ree+\agg\rgg)(2\chi\eta E'+E_2)+2\aeg\rge E_1^*]e^{-i\omega(t-z)}\nonumber\\
&+ [(\aee\ree+\agg\rgg)E_1^*+2\aeg^*\rge(E_2+2\chi\eta E')]e^{i\omega(t+z)}\nonumber\\
&+\left. [(\aee\ree+\agg\rgg)(2\chi\eta E'^*+E_2^*)+2\aeg^*\rge E_1]e^{i\omega(t-z)}\right\}\vec{\epsilon}'\nonumber\\
&+c.c.\,.
\end{align}
By matching the oscillation phases of the electric fields and the microscopic polarization and using the slowly varying envelope approximation, we arrive at the field equations for $E_1$, $E_2$ and $E'$
\begin{align}
(\partial_t-\partial_z)E_1&=\dfrac{i\omega n}{2}[(\aee\ree+\agg\rgg)E_1+2\aeg\rge^*(E_2^*+\chi\eta E'^*)]\,,\label{eq:eE1diff}\\
(\partial_t+\partial_z)E_2&=\dfrac{i\omega n}{2}[(\aee\ree+\agg\rgg)(E_2+\chi\eta E')+2\aeg\rge^*E_1^*]\,,\label{eq:eE2diff}\\
(\partial_t+\partial_z)E'&=\dfrac{i\omega^2 n}{\omega+k}[(\aee\ree+\agg\rgg)(2\chi^2\eta E'+\chi E_2)+2\aeg\rge^*\chi E_1^*]\,.\label{eq:eE'diff}
\end{align}
The first terms on the right hand sides of these equations, which are proportional to $\aee$ and $\agg$, do not affect the transition from excited to ground states, but rather describe absorption and reemission of photon or dark photons propagating in the medium. More importantly, the second terms on the right hand sides of the above equations, proportional to $\aeg$, describe the production of electromagnetic fields via excited to ground state transitions of the atoms. Altogether, $E_1$ can be amplified by seed $E_2$ and $E'$ fields, and correspondingly, $E_2$ and $E'$ are amplified by the $E_1$ field through transitions. For our purposes, we are most interested in the fact that $E'$ will amplify $E_1$ and $E_2$ in these equations, which forms the basis for our dark photon detection proposal.

\subsubsection{Bloch vector}
Defining the Bloch vector as in Section \ref{sec:2state}, from Eq.~\eqref{eq:eree} and Eq.~\eqref{eq:erge} we obtain
\begin{align}
&\partial_t r_1=\left[-\dfrac{\agg-\aee}{4}(|\bar{E}'_1|^2+|\bar{E}'_2|^2)+\delta\right]r_2+\aeg \im(\bar{E}'_1\bar{E}'_2)r_3-\dfrac{r_1}{T_2}\,,\label{eq:er1}\\
&\partial_t r_2=\left[\dfrac{\agg-\aee}{4}(|E'_1|^2+|E'_2|^2)-\delta\right]r_1+\aeg \re(\bar{E}'_1\bar{E}'_2)r_3-\dfrac{r_2}{T_2}\,,\label{eq:er2}\\
&\partial_t r_3=-\aeg[\im(\bar{E}'_1\bar{E}'_2)r_1+\re(\bar{E}'_1\bar{E}'_2)r_2]-\dfrac{1+r_3}{T_1}\,,\label{eq:er3}
\end{align}
where the spatially averaged visible and dark photon fields are together defined as
\begin{equation*}
\bar{E}'_1=\Eob+\chi\eta\Epb\,,\,\,\,\bar{E}'_2=\Etb+\chi\eta\Epb\,,
\end{equation*}
and we assume $\aeg$ is real. Note that the expression above has assumed that $\Eob$ and $\Etb$ are in phase, which is appropriate for atoms pumped by phase-matched lasers. Due to the smallness of the mixing parameter $\chi$, the dark photon field itself will not drive the evolution of the state population in the system. However, the dark photon can trigger the production of $E_1$ and $E_2$, which in turn trigger additional photon production. Therefore, while it would be safe to drop the dark photon component in \cref{eq:er1,eq:er2,eq:er3}, we retain it in numeric computations for the sake of rigor. 

Of course, because it is essential to the development of electromagnetic fields in $\ph$, we must retain the dark photon component in the field equations. Using \cref{eq:eE1diff,eq:eE2diff,eq:eE'diff} we obtain
\begin{align}
(\partial_t-\partial_z)E_1&=\dfrac{i\omega n}{2}\left[(\dfrac{\aee+\agg}{2}+\dfrac{\aee-\agg}{2}r_3)E_1+\aeg(r_1-ir_2)(E_2^*+\chi\eta E'^*)\right]\,,\label{eq:eE1bloch}\\
(\partial_t+\partial_z)E_2&=\dfrac{i\omega n}{2}\left[(\dfrac{\aee+\agg}{2}+\dfrac{\aee-\agg}{2}r_3)(E_2+\chi\eta E')+\aeg(r_1-ir_2)E_1^*\right]\,,\label{eq:eE2bloch}\\
(\partial_t+\partial_z)E'&=\dfrac{i\omega n}{2}\left[(\dfrac{\aee+\agg}{2}+\dfrac{\aee-\agg}{2}r_3)(2\chi^2\eta E'+\chi E_2)+\aeg(r_1-ir_2)\chi\eta E_1^*)\right]\,,\label{eq:eE'bloch}
\end{align}
In the experimental setup we soon describe, after the atoms are pumped into their excited states, the laser fields will be shut off so that $|\tilde{E}_1| = |\tilde{E}_2|\approx 0$. It is clear from Eq.~\eqref{eq:eE1bloch} that in this circumstance, a non-zero dark electric field $E'$ will be essential to develop the $E_1$ field, which will in turn trigger additional two photon emission.

\section{Detecting dark photon induced two photon transitions}
\label{sec:sens}

\subsection{Experimental setup}

\begin{figure}[!htb]
\centering
\includegraphics[width=0.8\textwidth]{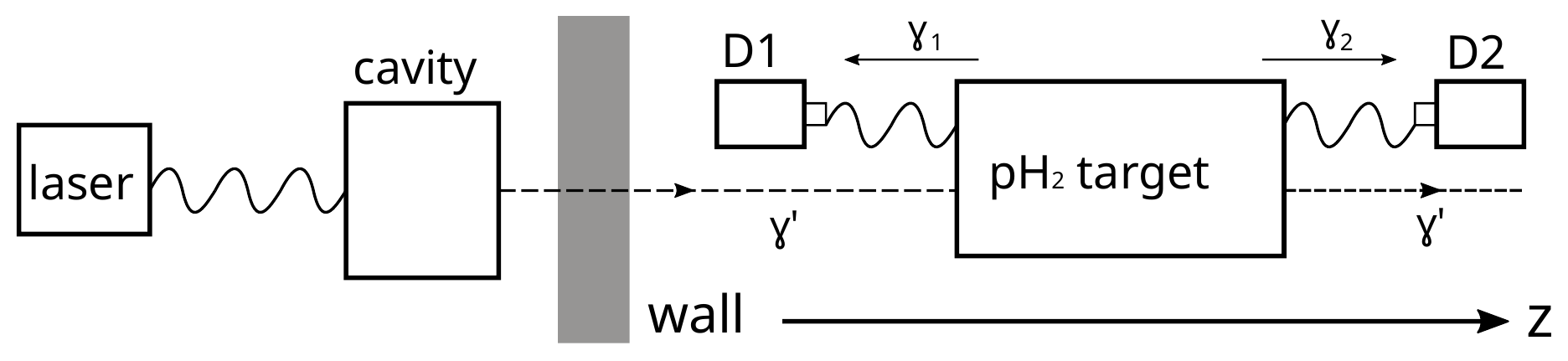}
\caption{Schematic view of the proposed experiment. First, the $\ph$ sample is coherently excited to energy $\weg$ by back-to-back pump lasers (pump lasers not shown). The excited atoms' $E1$ dipole transitions are parity forbidden, meaning the atoms are metastable over the $\sim$ ten nanosecond integration time of the experiment. On the other hand, the emission of two $\weg/2$ energy photons in an $E1 \times E1$ transition is allowed. As in light-shining-through wall experiments, a laser is fed into a resonant cavity to increase the dark photon conversion probability. In this case, the laser will operate at energy $\weg/2$, so that after passing through the wall, dark photons act as a trigger field for the emission of back-to-back photons which are then observed by detectors labeled D1 and D2.}
\label{fig:setup}
\end{figure}

Our proposed experimental setup is schematically illustrated in Figure~\ref{fig:setup}. A continuous laser beam is injected into a resonant cavity, which enhances the laser's probability to oscillate into dark photons. After hitting the wall photons are stopped and only dark photons are allowed through. The target ($\ph$ for example) is pumped into a coherently excited state as detailed in Section \ref{sec:2state}. As it propagates through the target, the dark photon field triggers atomic deexcitation. The electric fields generated from the first de-excitation subsequently triggers two photon emission, producing back-to-back photons with the same frequency. These photons trigger further deexcitations, detected at both ends of the target vessel.

There are two primary advantages to conducting the experiment in the manner described above. First, the $\ph$ sample's response to a dark photon field can be precisely determined by passing a very weak laser field through the sample, where low-power lasers can directly test the response to weakly coupled dark photons. Then, in discovery mode, where visible photons are prevented from passing through the wall, the two photon emission process would presumably only occur, if triggered by a dark photon over the $\sim$ 10 nanosecond coherence time, because the spontaneous deexcitation process is negligibly slow, as detailed in Section \ref{sec:spont}. Photons produced by dark photon induced transitions would be emitted back-to-back and at the frequency, $\omega = \weg /2$. Altogether this provides a powerful background rejection method, since the rate for spontaneous two photon emission is very small. This can be contrasted with more ambitious experiments utilizing two photon emission processes \cite{Fukumi:2012rn,Song:2015xaa,Yoshimura:2017ghb,Huang:2019rmc}. In these experiments, the signal (one photon and either two neutrinos or an axion) will need to be distinguished from a sizable two photon emission background, since both processes are triggered. Therefore it is plausible that the experiment we have outlined is an intermediate step that could be reached while working towards the proposals laid out in \cite{Fukumi:2012rn,Song:2015xaa,Yoshimura:2017ghb,Huang:2019rmc}.

\subsection{Dark photon induced transition rate}
\label{sec:dpinducerate}
To begin with, let us quote the estimated rate for the emission of $\gamma_1$ and $\gamma_2$ in our proposed experiment. First, we note that without both coherent enhancement and exponential amplification of photon fields by the atomic medium that will be discussed shortly, the $| e \rangle \rightarrow | g \rangle + \gamma + \gamma'$ transition rate depicted in Figure~\ref{fig:transit_a} is rather slow. To satisfy the coherent amplification condition, we require
\begin{equation}
(\omega-k)L\lesssim 1\,,
\label{eq:cohamp}
\end{equation}
where $L$ is the length of the target, which is the longest dimension of the target volume. 

Under these conditions (see Appendix \ref{app:base} for a full derivation) the naive rate for dark photon-induced two stage transitions is
\begin{equation}
\Gamma_\mathrm{\gamma'\gamma}=\dfrac{1}{4\pi}(N_\mathrm{pass}+1)\chi^4\sin^2\left(\dfrac{\mAp^2}{4\omega'}l\right)P_L|\eta|^2|\aeg|^2|\rge|^2n^2V^2\omega_1^3\,,
\end{equation}
where $\omega_1 $ is the cavity laser frequency, equal to the dark photon frequency $\omega'$, $N_\mathrm{pass}$ is the number of cavity reflections, $P_L$ is the cavity laser power, $l$ is the cavity length, $A$ is the area of the excited atomic target (limited by the pump lasers' beam width) and $n$ is the target number density. In Table \ref{tab:bench}, we give the parameters for the laser cavity and parahydrogen sample in our benchmark setup.

\begin{table}[h!]
\centering
 \begin{tabular}{||c|| c||} 
 \hline
 Dark Photon Generating Cavity & Superradiant Parahydrogen Target  \\ [0.5ex] 
 \hline\hline
Cavity Length  $l=50\ \mathrm{cm}$ & Sample Length  $L=30\ \mathrm{cm}$  \\ [.1ex] \hline
Cavity Reflections $N_\mathrm{pass}=2\times 10^4$ & pH$_2$ Density $n =10^{21} ~{\rm cm^{-3}}$ \\  [.1ex]\hline
 Cavity Laser Freq. $\omega'=0.26~{\rm eV} $ & Pump Laser Freq. $\omega_1=0.26~{\rm eV} $   \\ [.1ex] \hline
Cavity Laser Power $P_L = 1~{\rm W~mm^{-2}}$ & Pump Laser Power $ \approx 10^9~{\rm W~mm^{-2}}$  \\ [.1ex] \hline
-- & pH$_2$ Sample Area $A = 1~{\rm cm^2}$  \\ [.1ex]
 \hline
 \end{tabular}
 \caption{Parameters are given for our benchmark experimental setup. For the dark photon generating laser cavity, we take parameters matching those of the ALPS experiment \cite{Bahre:2013ywa}. For the pH$_2$ sample, we quote values necessary to obtain maximum coherence, as investigated in Section \ref{sec:2state}.}
 \label{tab:bench}
\end{table}

Using this naive estimate results in an unobservably small rate, because it does not account for the development of electromagnetic fields in the atomic medium ($c.f.$ the field equations given in Eqs.~\eqref{eq:eE1bloch} through \eqref{eq:eE'bloch}). The predicted rate for our benchmark experimental and model parameters given in Table \ref{tab:bench}, for a dark photon mixing  $\chi=10^{-9}$ and mass $\mAp=10^{-4}\ \mathrm{eV}$, and for parahydrogen dipole coupling $\aeg=0.0275\times10^{-24}\ \mathrm{cm}^3$, is 
$
\Gamma \approx 10^{-5}\ \mathrm{s}^{-1}\,.
$
This emission rate is unobservably low considering that each experimental run is expected to last about $10\ \mathrm{ns}$. 

However, even a small production rate for $E_1$ can be exponentially enhanced in coherently prepared atoms. As detailed in Appendix \ref{sec:tpeqed}, the transition rate for producing two photon pairs is exponentially enhanced as the electromagnetic field strength grows,
\begin{equation}
\Gamma_\mathrm{sup}=\dfrac{1}{16\pi}|\aeg|^2|\rge|^2N^2V\omega_1^2|E_1|^2|E_2|^2\,.
\label{eq:tperate}
\end{equation}
The dependence on $E_1^2E_2^2$ in Eq.~\eqref{eq:tperate} shows that the growth of signal fields will be exponential after the dark photon establishes a small $E_1$ seed field. A similar amplification has been observed to be as large as $10^{18}$ compared with spontaneous emission~\cite{Miyamoto:2015tva}. We expect an even larger amplification factor to be achieved in our benchmark experimental setup.

\subsection{Numerically simulating field development}
\label{sec:numerics}
When simulating the development of electric fields in coherently prepared atoms, it will be convenient to rescale the spacetime coordinates and the electric fields to be dimensionless. We define
\begin{equation}
\beta=\dfrac{2}{n\weg \aeg}\,,\,\,\, \zeta=\dfrac{z}{\beta}\,,\,\,\,\tau=\dfrac{t}{\beta}\,,\,\,\,|e_{1(2)}|^2=\dfrac{|E_{1(2)}|^2}{\weg n}\,,\,\,\,|e'|^2=\dfrac{|E'|^2}{\weg n}\,,
\end{equation}
where $\beta$ represents the typical length and time scale for the evolution of the system and $\weg n$ is the energy stored in excited atoms. The Bloch equations and field equations can be written in terms of these new variables
\begin{align}
\partial_\tau r_1&=\left[-\dfrac{\agg-\aee}{2\aeg}(|e_1|^2+|e_2|^2)+\beta\delta\right]r_2+2\im(e_1e_2)r_3-\dfrac{r_1}{\tau_2}\,,\label{eq:er1dim}\\
\partial_\tau r_2&=\left[\dfrac{\agg-\aee}{2\aeg}(|e_1|^2+|e_2|^2)-\beta\delta\right]r_1+2 \re(e_1e_2)r_3-\dfrac{r_2}{\tau_2}\,,\label{eq:er2dim}\\
\partial_\tau r_3&=-2[\im(e_1e_2)r_1+\re(e_1e_2)r_2]-\dfrac{1+r_3}{\tau_1}\,,\\
(\partial_\tau-\partial_\zeta)e_1&=\dfrac{i}{2}\left[(\dfrac{\aee+\agg}{2\aeg}+\dfrac{\aee-\agg}{2\aeg}r_3)e_1+(r_1-ir_2)(e_2^*+\chi\eta e'^*)\right]\,,\label{eq:eE1dim}\\
(\partial_\tau+\partial_\zeta)e_2&=\dfrac{i}{2}\left[(\dfrac{\aee+\agg}{2\aeg}+\dfrac{\aee-\agg}{2\aeg}r_3)(e_2+\chi\eta e')+(r_1-ir_2)e_1^*\right]\,,\label{eq:eE2dim}\\
(\partial_\tau+\partial_\zeta)e'&=\dfrac{i}{2}\left[(\dfrac{\aee+\agg}{2\aeg}+\dfrac{\aee-\agg}{2\aeg}r_3)(2\chi^2\eta e'+\chi e_2)+(r_1-ir_2)\chi\eta e_1^*)\right]\,,\label{eq:eE'dim}
\end{align}

As mentioned before, dipole couplings of parahydrogen have been measured to be $\agg=0.90 \times 10^{-24}\ \mathrm{cm}^3$, $\aee=0.87 \times 10^{-24}\ \mathrm{cm}^3$, $\aeg=0.0275 \times 10^{-24} \mathrm{cm}^3$ \cite{Yoshimura:2012tm}. For the relaxation and decoherence times, we take $T_1=10^3\ \mathrm{ns}$ and $T_2=10\ \mathrm{ns}$ respectively; for extended discussion of coherence in preparations of $\ph$, see Section \ref{sec:2state}. The photon and dark photon energies are $\omega=\weg/2 \approx 0.26\ $eV. Altogether this gives
\begin{align}
&\beta= 0.092\left(\dfrac{10^{21}~\mathrm{cm}^{-3}}{n}\right)\mathrm{ns}= 2.8\left(\dfrac{10^{21}~\mathrm{cm}^{-3}}{n}\right)\mathrm{cm}\,,\nonumber\\
&\weg n=2.5*10^{10}\left(\dfrac{n}{10^{21}~\mathrm{cm}^{-3}}\right)\mathrm{W/}\mathrm{mm}^2\,.
\end{align}
A typical target vessel is 10 to 100~cm long. Here we assume a vessel that is 30~cm long, which is smaller than the expected length scale over which the $\ph$ is coherent. If we assume all atoms are initially prepared in the coherent state, then $r_1=1$ across the target. We will also consider smaller values of $r_1 =0.1,0.5,0.9$, which correspond to fewer atoms in the coherent state. With the aid of a resonant cavity, the transmission probability for a dark photon to shine through the wall is given by~\cite{Ahlers:2007rd,Ahlers:2007qf}
 \begin{equation}
 p_\mathrm{trans}=2 (N_\mathrm{pass}+1) \chi^2\sin^2\left(\dfrac{\mAp^2}{4\omega}l\right)\,,
 \end{equation}
where $l$ is the size of the cavity and $N_\mathrm{pass}$ is the number of reflections the laser undergoes in the dark photon generating cavity. We assume the laser cavity has parameters given in Table \ref{tab:bench}; these values are in line with what has been attained at the ALPS II experiment \cite{Bahre:2013ywa}. The initial dark photon field power in the target volume is estimated to be
\begin{equation}
|E'(t=0)|^2 = P_\mathrm{L}\ p_\mathrm{trans}\,,
\label{eq:eprime0}
\end{equation}
where for our benchmark setup we assume a laser power $P_\mathrm{L}=1\ \mathrm{Wmm}^{-2}$. As mentioned in Section~\ref{sec:maxwelleq} $\eta$ is determined by the relative phase between the polarization of the photon and dark photon fields. Without loss of generality we set it to be unity here.

\begin{figure}[!htb]
\centering
\begin{subfigure}{0.49\textwidth}
\includegraphics[width=\linewidth]{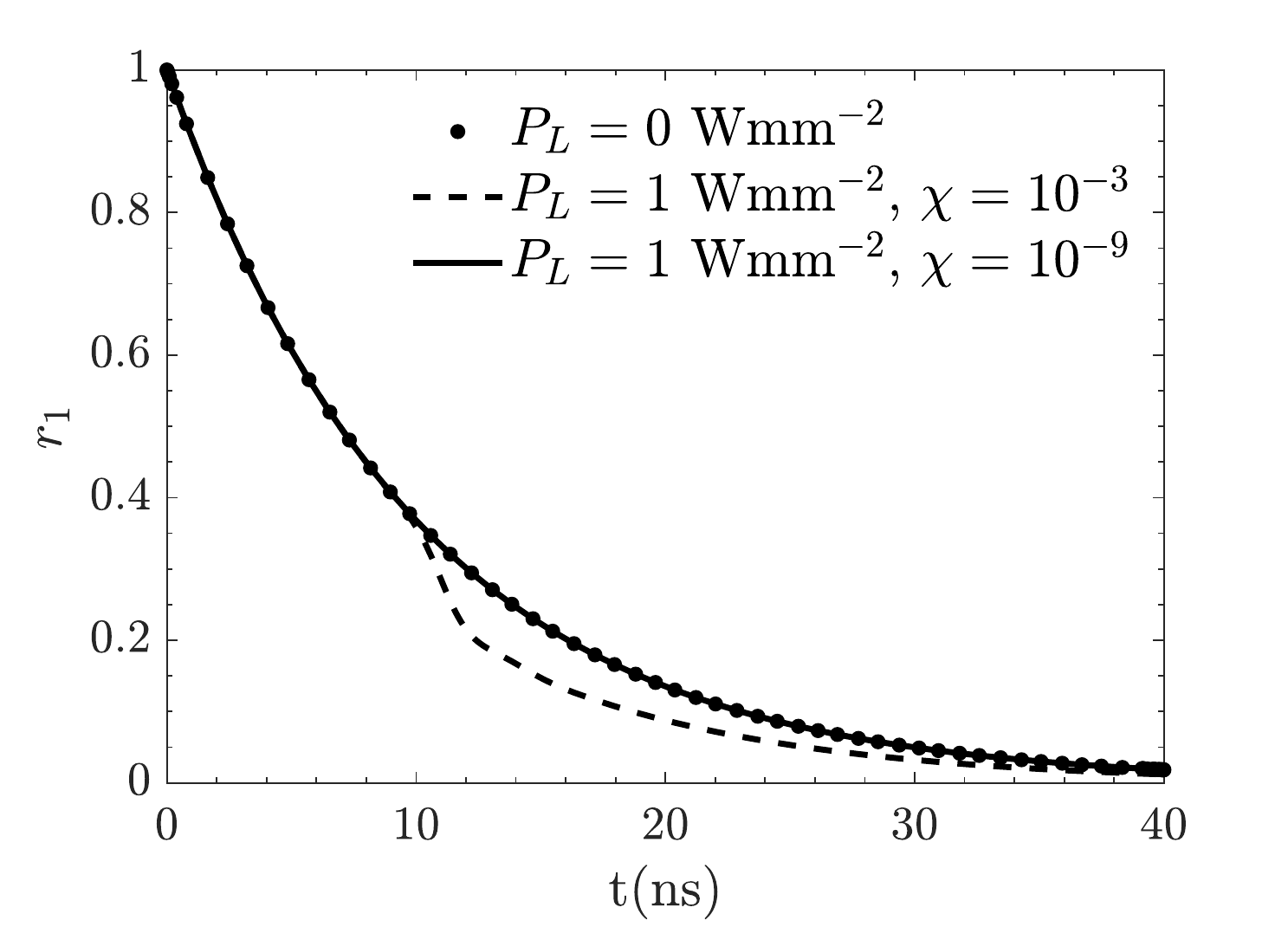}
\end{subfigure}
\hspace*{-0.15cm} 
\begin{subfigure}{0.49\textwidth}
\includegraphics[width=\linewidth]{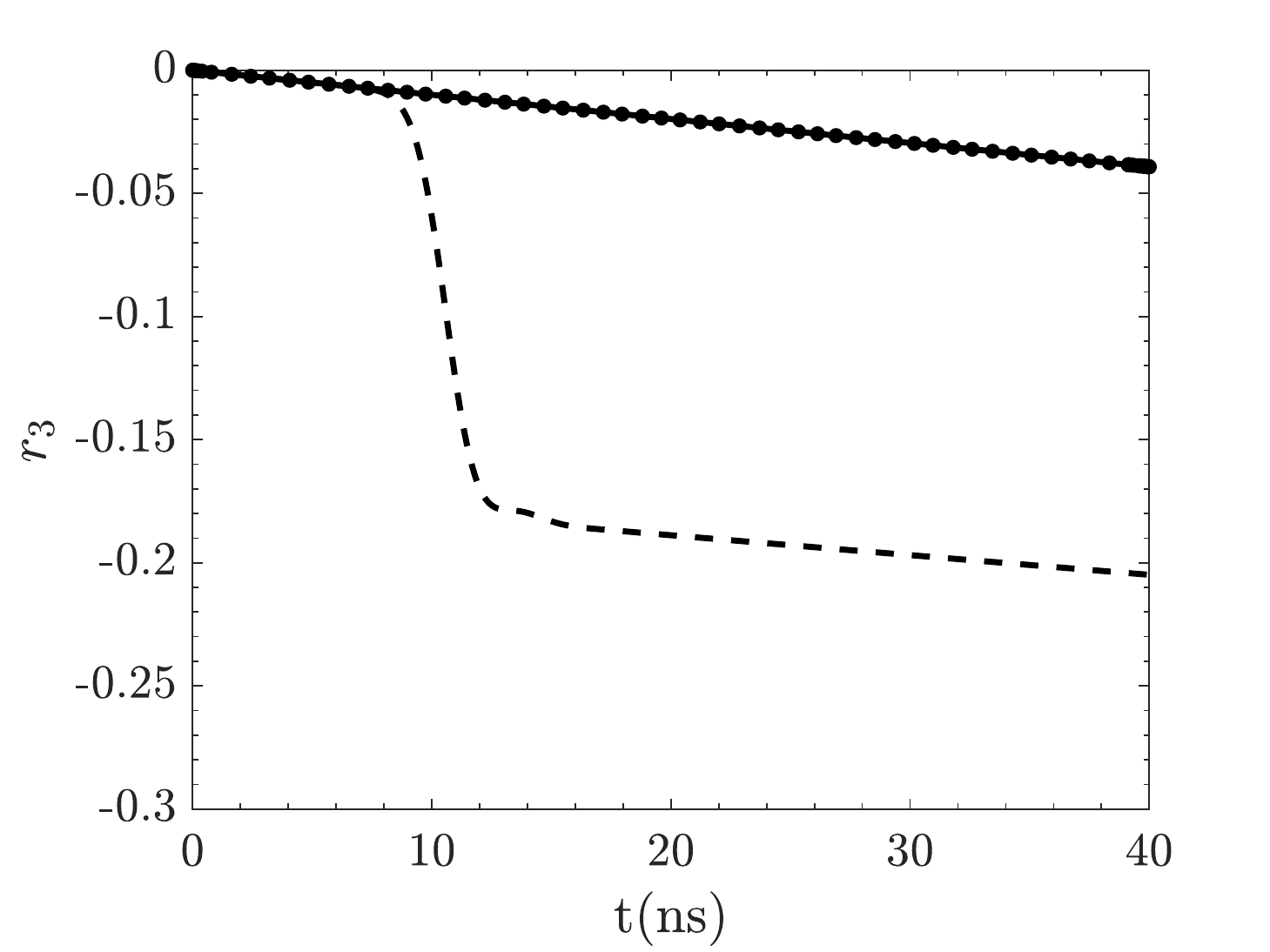}
\end{subfigure}
\caption{Time evolution of Bloch vectors at the center of the target. Left panel: $r_1$ as a function of time. Right panel: $r_3$ as a function of time. Dotted lines show the case where no initial cavity laser is present to create a dark photon ($P_L = 0$). The dashed and solid lines corresponds to dark photon mixing parameters $\chi=10^{-3}$ and $10^{-9}$ respectively, for a cavity laser generating dark photons with power $P_L = 1\ \mathrm{Wmm}^{-2}$. A maximally coherent parahydrogen sample, $r_1(t=0)=1$, $r_2(t=0)=0$, $r_3(t=0)=0$ and dark photon mass $\mAp=0.1\ $meV are assumed.}
\label{fig:blochvector}
\end{figure}

\begin{figure}[!htb]
\centering
\begin{subfigure}{0.49\textwidth}
\includegraphics[width=\linewidth]{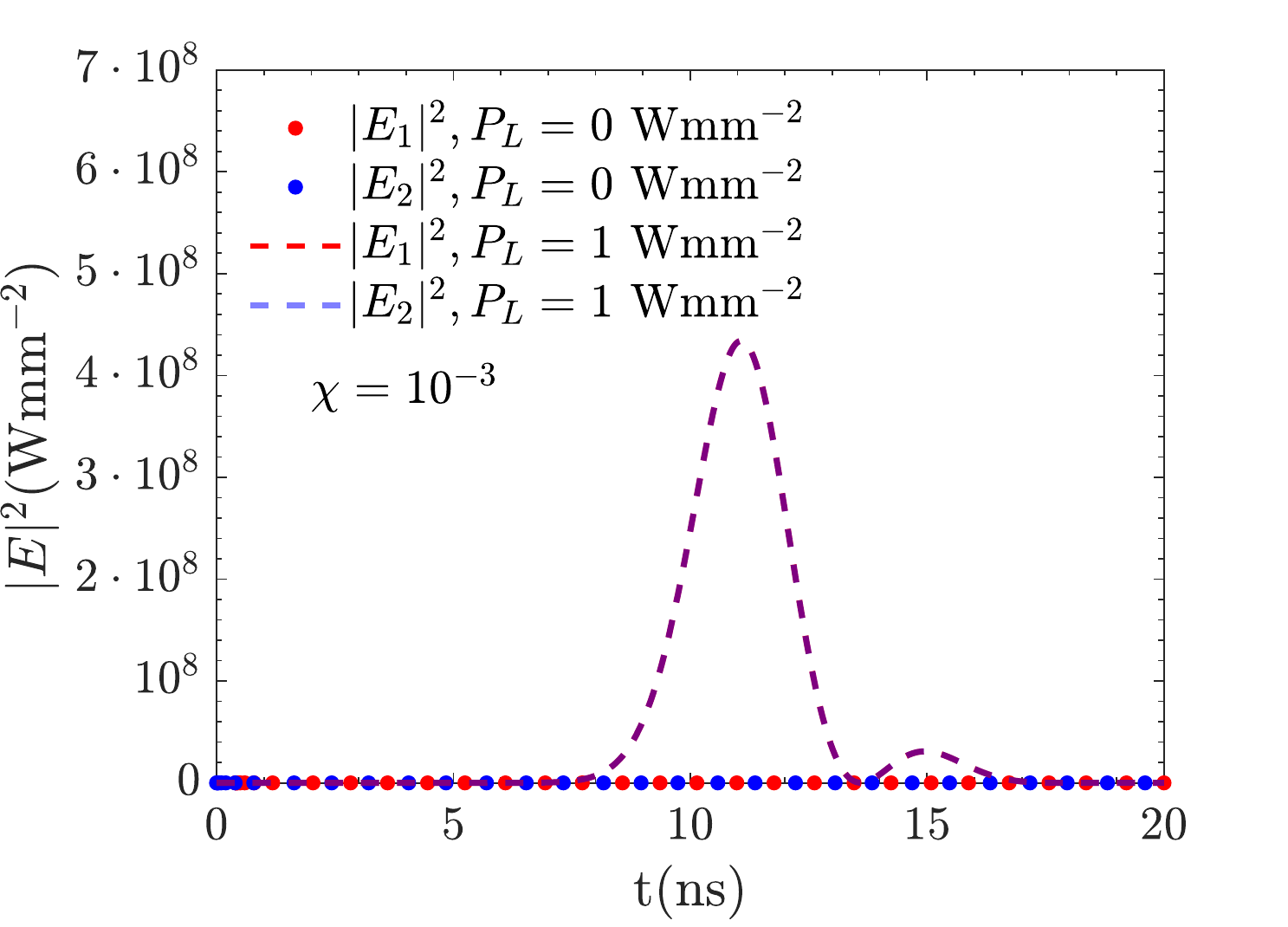}
\end{subfigure}
\hspace*{-0.15cm} 
\begin{subfigure}{0.49\textwidth}
\includegraphics[width=\linewidth]{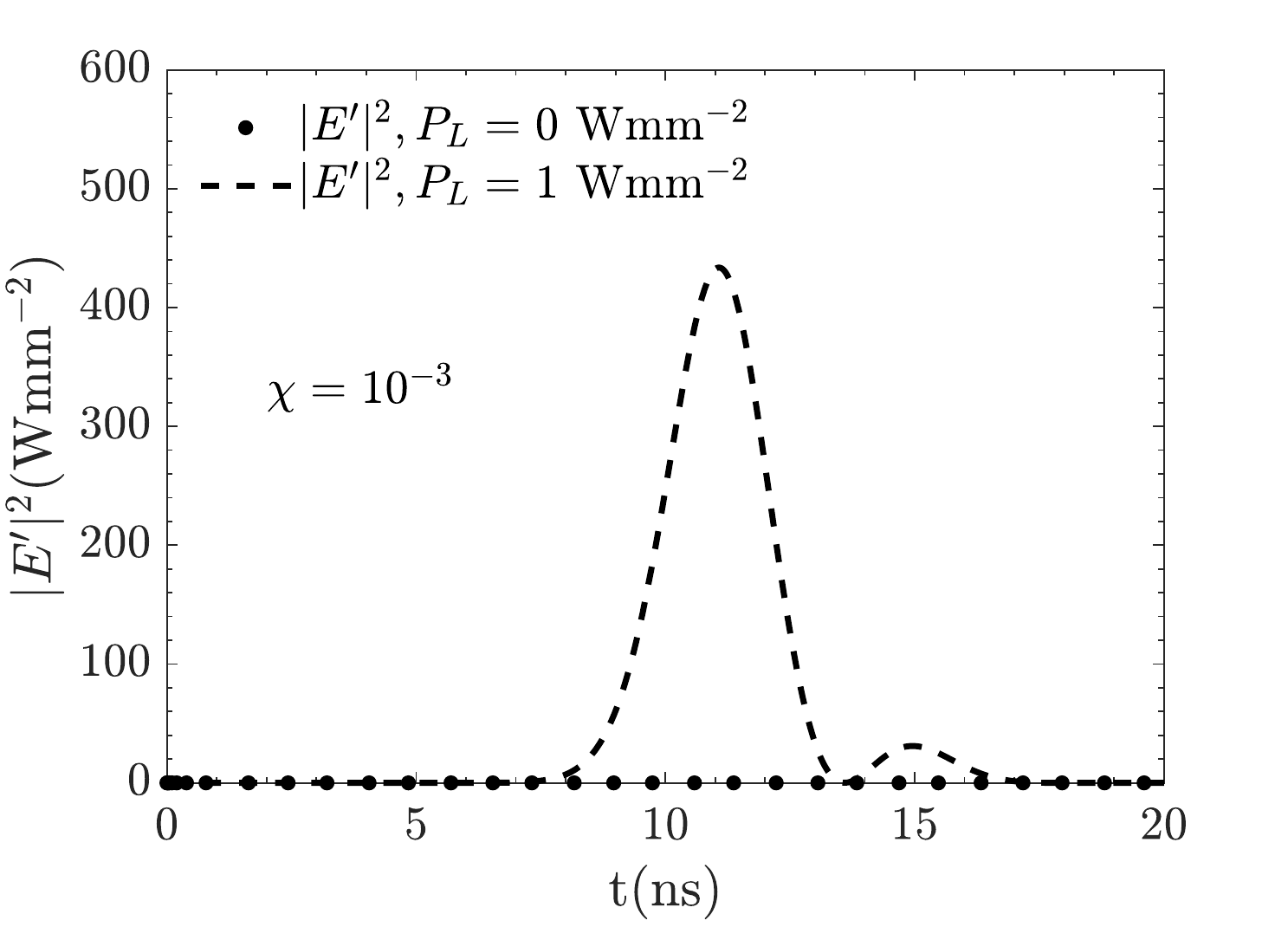}
\end{subfigure}
\caption{Time evolution of the (dark) electric fields at the ends of the parahydrogen target. Left panel: The electric fields at each end of the target, $|E_1|^2$ and $|E_2|^2$, as a function of time. Right panel: the dark photon field $|E'|^2$ as a function of time. As in Figure~\ref{fig:blochvector} the cavity laser is turned off ($P_L=0$) for the dotted lines and a $P_L = 1\ \mathrm{Wmm}^{-2}$ cavity laser with dark photon mixing $\chi=10^{-3}$ is assumed for the dashed lines. We also assume the same initial Bloch vectors as Figure~\ref{fig:blochvector}. $|E_1|^2$ (red) is taken at the left end of the target with $z=0\ \mathrm{cm}$ while $|E_2|^2$ (blue) and $|E'|^2$ (black) are taken at the right end of the target with $z=30\ \mathrm{cm}$.}
\label{fig:fields_1e-3}
\end{figure}

\begin{figure}[!htb]
\centering
\begin{subfigure}{0.49\textwidth}
\includegraphics[width=\linewidth]{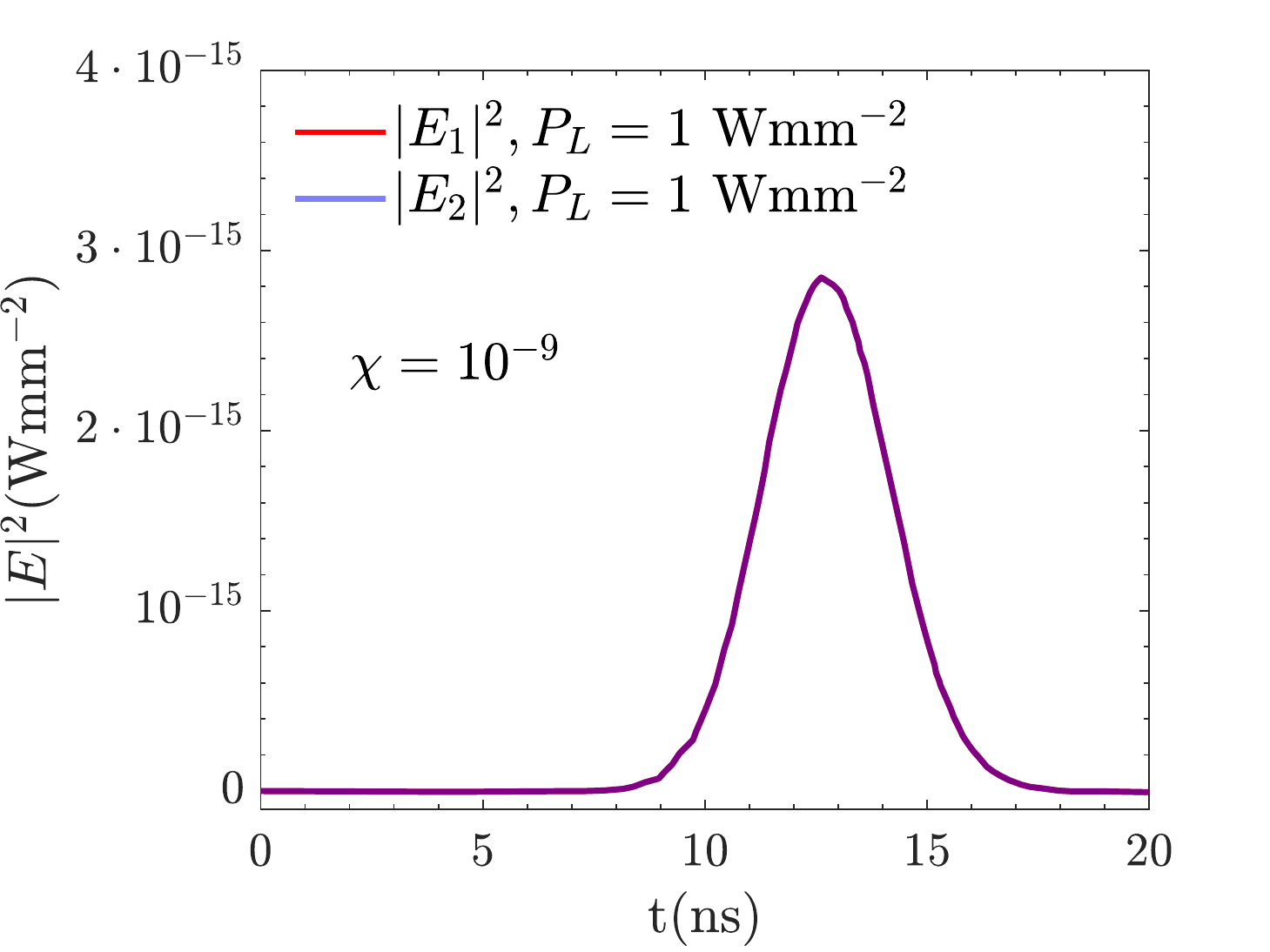}
\end{subfigure}
\hspace*{-0.15cm} 
\begin{subfigure}{0.49\textwidth}
\includegraphics[width=\linewidth]{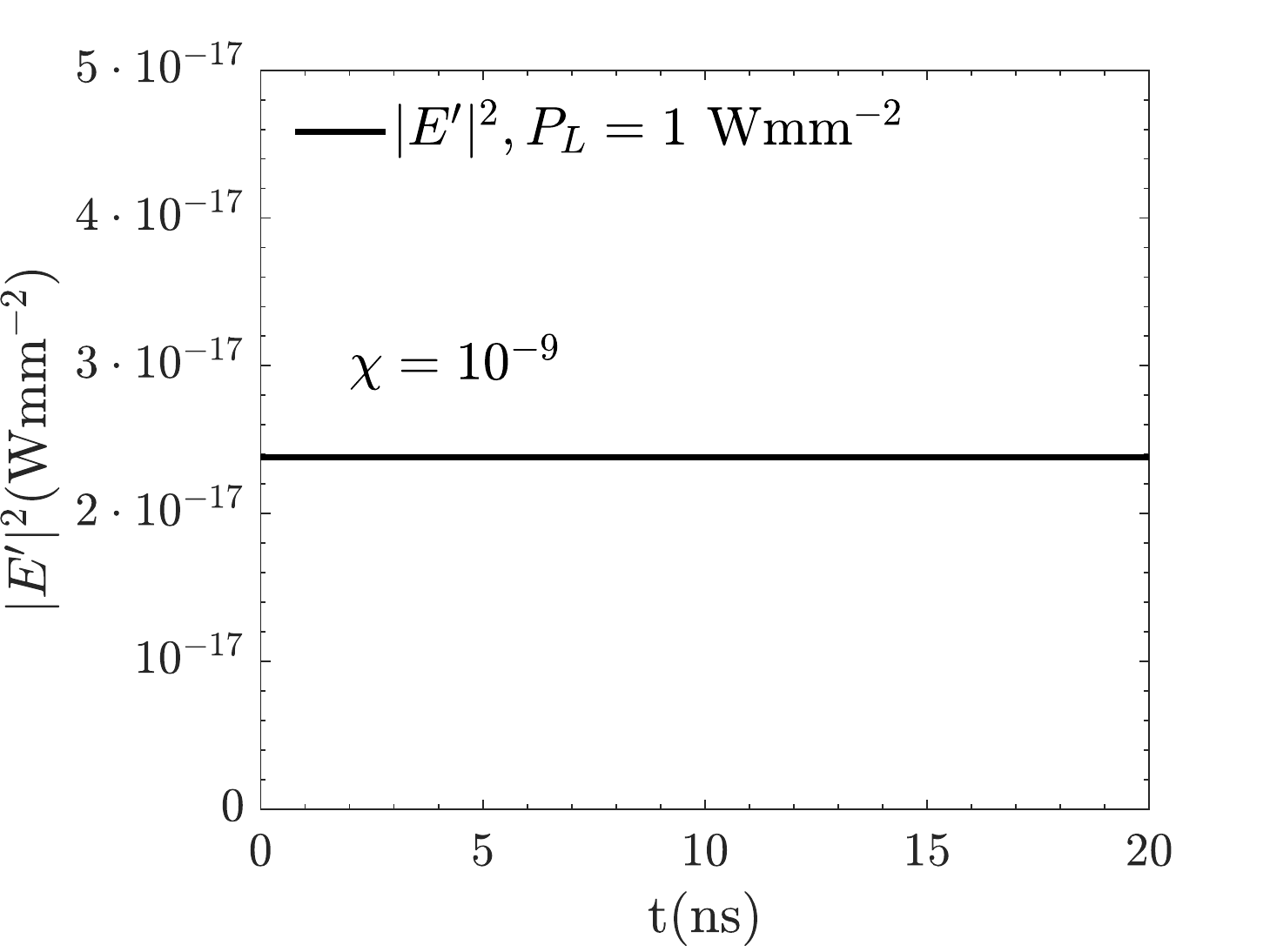}
\end{subfigure}
\caption{Time evolution of the (dark) electric fields at the ends of the target. Same as Figure~\ref{fig:fields_1e-3} but a smaller dark photon mixing $\chi=10^{-9}$ is used and $|E_1|^2$, $|E_2|^2$ and $|E'|^2$ are shown with solid lines.}
\label{fig:fields_1e-9}
\end{figure}

We show in Figure~\ref{fig:blochvector} through Figure~\ref{fig:fields_1e-9} the time evolution of the system. In these figures we assume all the $\ph$ atoms are initially prepared in the coherent state, $i.e.$ $r_1=1$, $r_2=0$ and $r_3=0$, across the target. We also assume a dark photon mass $\mAp=0.1\ $meV.

As shown in Figure~\ref{fig:blochvector} $r_1$ and $r_3$ decay exponentially when no laser is present. In this case no initial dark photon field is pumped through the wall and so spontaneous deexcitation dominates the evolution of the system. We note that Figure~\ref{fig:fields_1e-3}, which shows no substantial $E_1$, $E_2$, or $E'$ field developing when $P_L=0$, has not included the effect of spontaneous two photon deexcitations, which are expected to be negligibly small, see Section \ref{sec:spont}.

This scenario changes dramatically in the presence of dark photons produced by a laser. Assuming the laser power $P_L=1\ \mathrm{Wmm}^{-2}$ and the mixing $\chi=10^{-3}$, a sudden drop takes place in $r_1$ and $r_3$ around 10 ns. This drop corresponds to decay and release of the target's energy through production of $E_1$ and $E_2$ as well as a minor enhancement of the dark photon field $E'$. The dynamics can be explained as follows. The initial dark photon fields induces a deexcitation via $E_1$ and $E'$ (Figure~\ref{fig:transit_a} illustrated this process), then this $E_1$ field triggers additional two photon deexcitation producing $E_1$ and $E_2$ symmetrically (see Figure~\ref{fig:transit_b}). The growing $E_1$ and $E_2$, when large enough, cause abrupt decoherence and deexcitation, which in turn gives rise to additional energy release in the form of $E_1$ and $E_2$. As can be identified from Eq.~\eqref{eq:eE'diff} $E'$ will also be generated by $E_1$ induced transitions, at a rate suppressed by $\chi$.

The transitions are less explosive when $\chi=10^{-9}$, as illustrated in Figure~\ref{fig:blochvector} and Figure~\ref{fig:fields_1e-9}. The deviations of $r_1$ and $r_3$ from spontaneous decay are barely observed and the peak intensity of $E_1$ and $E_2$ are relatively low compared to $\chi=10^{-3}$. In this case, the dark photon has induced the generation of an observable but small quantity of $E_1$ and $E_2$ photons. The dark photon field remains essentially constant since $E'$ regenerated from $E_1$ is too small to be observed.

\subsection{Spontaneous two photon emission background}
\label{sec:spont}
We now consider a possible background from spontaneous deexcitation and emission of photons from cold atoms over the runtime of the proposed experiment (around 10 ns). We will find that this background is negligible. Since the transition from excitation state $|e\rangle$ to the ground state $|g\rangle$ is $E1$ forbidden, single photon deexcitation is only viable through higher order transitions. Note that we are only looking for signal photons with energy around $\omega=\frac{1}{2}\weg$, because our signal photons are expected at this frequency. The background from spontaneous two photon emission has a rate given by (see Appendix \ref{sec:tpeqed})
\begin{equation}
\dfrac{d\Gamma_\mathrm{sp}}{dz}=\dfrac{\weg^7}{(2\pi)^3}N|\aeg|^2z^3(1-z)^3=1.27\times 10^{-14}\ \mathrm{s}^{-1}\,,
\end{equation}
where $z=\omega_1/\weg$ is the fraction of the energy for one of the two photons in the transition. We assume an uncertainty $\Delta \nu=100\ $MHz in the frequency measurement, which translates to $\Delta z=8.0\times 10^{-7}$. For a sample target with length $L=30\ $cm and cross section area $A=1\ $cm$^2$ the uncertainty in the emission solid angle is $\Delta \Omega/4\pi=A/4\pi(L/2)^2=3.5\times 10^{-4}$. These two photons from spontaneous decay process can be emitted in any direction. Since we only detect photons at the ends of the atomic sample, the fraction of background photons which reach the detector is $2\Delta \Omega/4\pi$. Given the target number density $n=10^{21}\ $cm$^{-3}$ and complete coherence ($\reg=0.5$) the total number of pH$_2$ atoms in the excitation state for our benchmark setup is $N=1.5\times10^{22}$. Even given a generously long measurement time $\Delta t=40\ $ns, the spontaneous two photon emission background is estimated to be
\begin{equation}
N_{background}=2 N\dfrac{d\Gamma_\mathrm{sp}}{dz}\Delta z\Delta t \dfrac{\Delta \Omega}{4\pi} =4.3\times 10^{-9}\,.
\end{equation}
We see that over the course of any reasonable number of experimental repetitions, we should not expect a single background event from spontaneous two photon deexcitation processes.

\subsection{Results and sensitivity}
\begin{figure}[t!]
\centering
\includegraphics[width=0.6\textwidth]{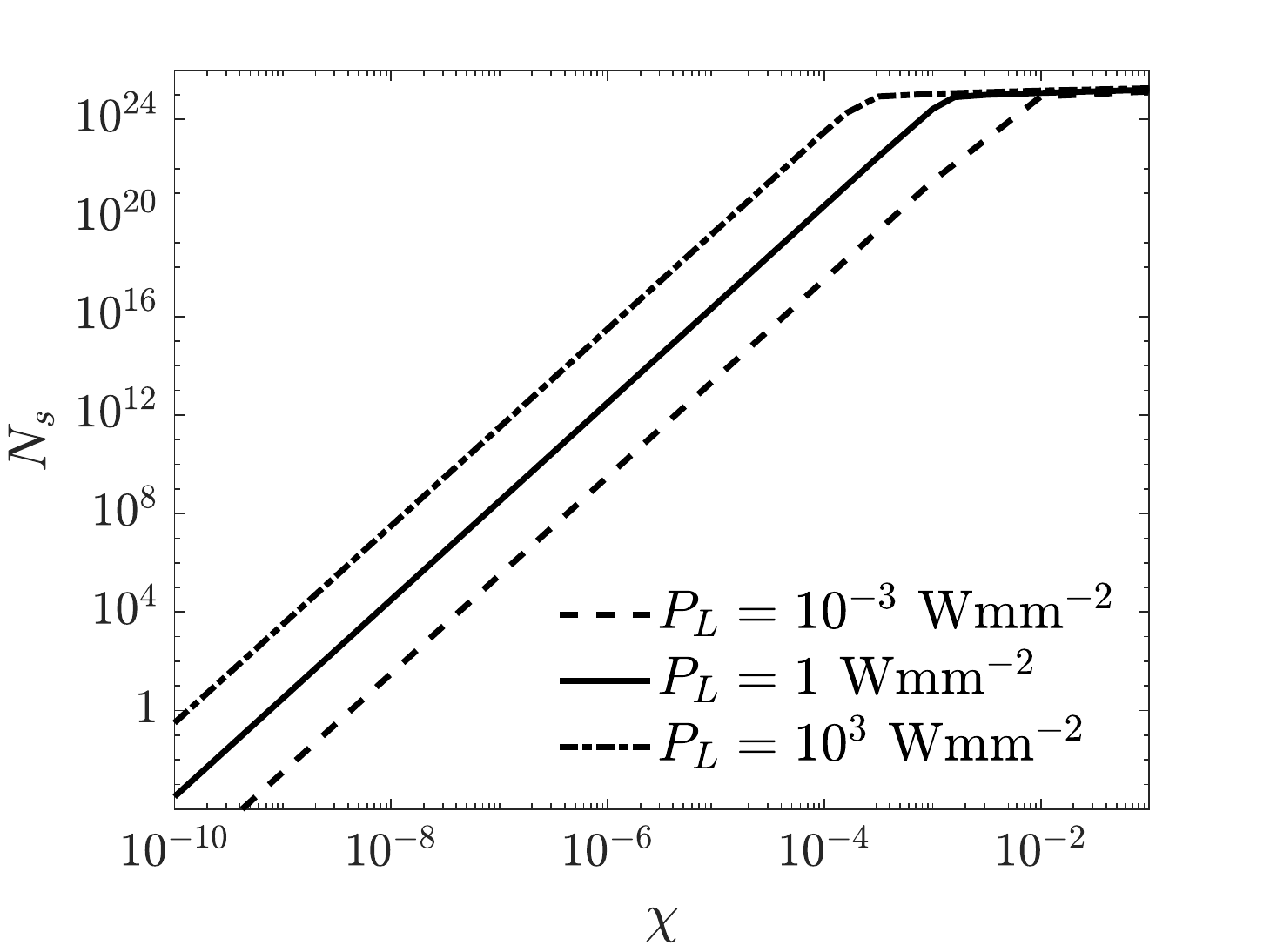}
\caption{Number of photons emitted as a function of $\chi$ at the ends of the target volume, for different $P_L$, where this is the power of the laser producing dark photons in the reflection cavity. The dashed, solid and dash dotted lines correspond to dark photon cavity laser powers $P_L=10^{-3},\,1,\,10^3\ \mathrm{Wmm}^{-2}$ respectively. Otherwise, we take the benchmark parameters shown in Table \ref{tab:bench}; the number of cavity reflections is $N_\mathrm{pass}=2\times 10^4$, the cavity length is $l=50\ \mathrm{cm}$, the dark photon mass is $\mAp=0.1\ \mathrm{meV}$, the laser frequency $\omega=0.26\ \mathrm{eV}$, and the area of the parahydrogen target is $A=1\ \mathrm{cm}^2$, and the number of experimental repetitions (each around 10 ns) is $N_\mathrm{rep}=10^3$. The initial Bloch vectors are the same as Figure~\ref{fig:blochvector}.}
\label{fig:Nsofchi}
\end{figure}

\begin{figure}[ht!]
\centering
\includegraphics[width=0.85\textwidth]{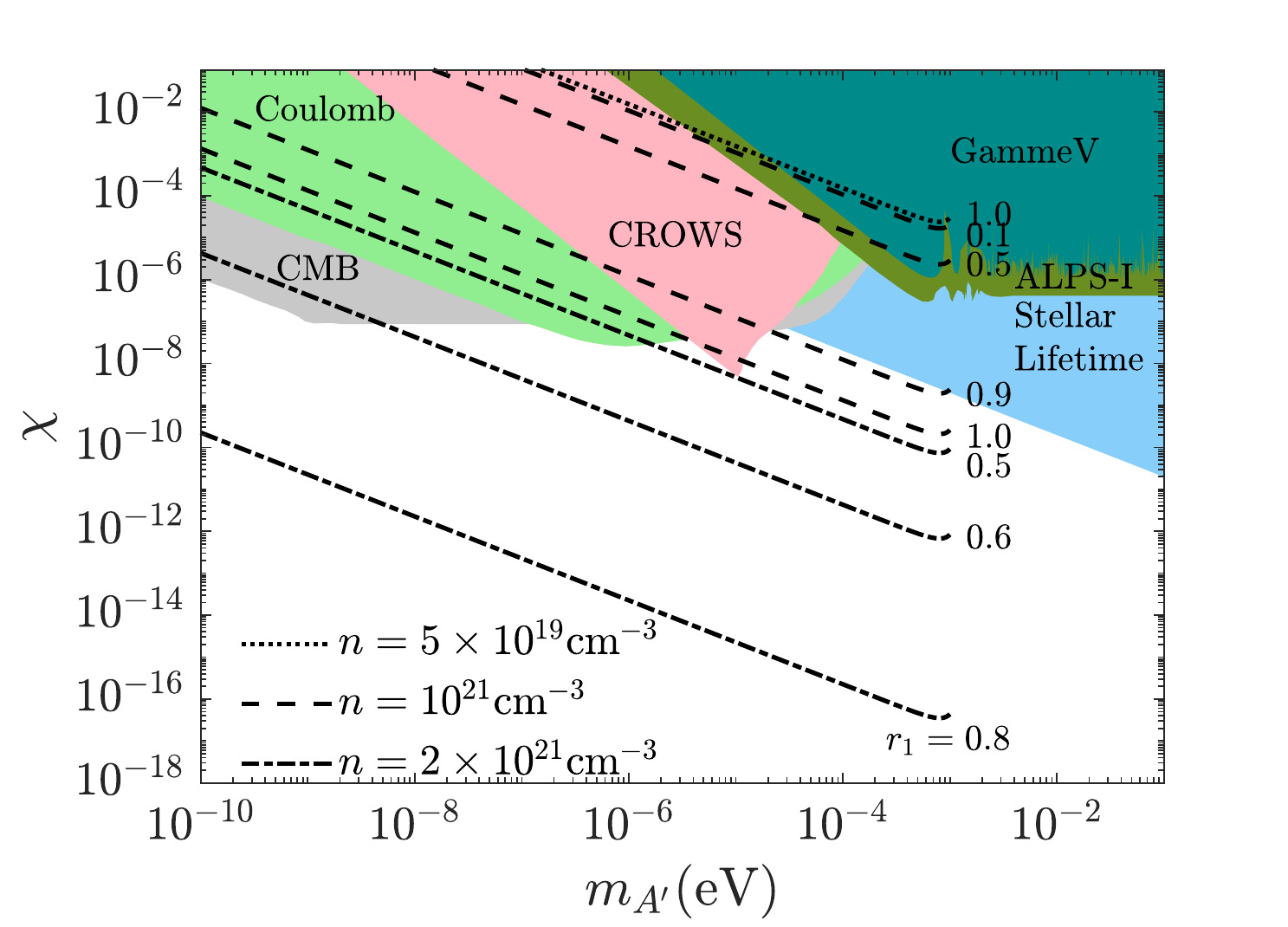}
\caption{The sensitivity of our proposed experiment assuming the benchmark parameters given in Table \ref{tab:bench}, except for $\ph$ target number density $n$ and Bloch vector $r_1(t=0)$ as indicated. We also assume $r_2(t=0)=r_3(t=0)=0$. The experiment is repeated $N_\mathrm{rep}=10^3$ times, by coherently exciting the $\ph$ sample, where each coherent excitation lasts $\sim$ 10 ns. The constraints from other dark photon experiments, astrophysics, and cosmology are shown for comparison, $e.g.$ Coulomb \cite{Jaeckel:2010xx,Williams:1971ms}, CMB \cite{Jaeckel:2008fi,Mirizzi:2009iz}, CROWS \cite{Betz:2013dza}, GammeV \cite{Steffen:2009sc}, ALPS \cite{Ehret:2010mh,Bahre:2013ywa}, and stellar constraints \cite{An:2013yfc,An:2013yua,Bahre:2013ywa,Hardy:2016kme}, see \cite{Redondo:2015iea} for a summary of these bounds. The black lines show the sensitivity of our proposed experiment, for $\ph$ number densities indicated, and coherence factors ($r_1$) as indicated to the right of each sensitivity. Section \ref{sec:2state} and particularly Table \ref{tab:decoh} provide a discussion of the coherence that has been achieved in ongoing $\ph$ experiments \cite{Hiraki:2018jwu} -- the maximum coherence factor so far obtained is $r_1 \sim 0.06$. The experimental sensitivities shown are obtained by numerically solving Eqs.~\eqref{eq:er1} through \eqref{eq:eE'bloch}, which determine the nonlinear development of coherence induced by counter-propagating lasers in $\ph$. As has been noted in past literature, the $\ph$ transition rate and the resulting experimental sensitivity exhibit dramatic nonlinear sensitivity to the $\ph$ coherence \cite{Yoshimura:2006nd,Yoshimura:2008ya,Fukumi:2012rn,Miyamoto:2015tva,Hara:2017,Hiraki:2018jwu}. This accounts for the dramatic increase in sensitivity between the bottom three sensitivity curves for coherence values $r_1 = 0.5 - 0.8$. The sensitivity curves have been truncated at a dark photon mass $m_{A'} = {\rm meV}$, beyond which the coherent amplification condition, Eq.~\eqref{eq:cohamp}, is no longer valid.}
\label{fig:sensitivity}
\end{figure}

The signature of the proposed dark photon search is the symmetric emission of photons with frequency $\omega=\weg/2$ at both ends of the target. The number of signal photons emitted during one experimental trial run (of $\sim 10$ ns) is
\begin{equation}
N_s^1=\dfrac{A}{\omega}\int_0^t |E_1(t')|^2dt'=\dfrac{A}{\omega}\int_0^t |E_2(t')|^2dt'\,,
\end{equation}
where $A$ is the area of the target and $t$ is the time duration of the experiment. The experiment can be repeated many times to accumulate signal photons. The Bloch equations and field equations derived in Section~\ref{sec:darkSR} are highly nonlinear, but we see from Eq.~\eqref{eq:eE1bloch} that $E_1\propto \chi E'$, and therefore the number of photons emitted is
\begin{equation}
N_s\propto P_L N_\mathrm{rep}\chi^4(N_\mathrm{pass}+1)\sin^2\left(\dfrac{\mAp^2}{4\omega}l\right)\,,
\label{eq:Nsrelation}
\end{equation}
where $N_\mathrm{rep}$ is the number of repetitions of the experiment.  To see in what regime this scaling holds, we show in Figure~\ref{fig:Nsofchi} the number of photons produced as a function of the mixing, $\chi$ assuming different laser powers. Note that in the limit that $\mAp^2 \ll \frac{4 \omega}{l} $, the number of signal photons expected scales like $N_s \propto \chi^4 \mAp^4$, as is evident in Figures \ref{fig:Nsofchi}. There is an upper bound on the number of signal photons, which is saturated if all of the excited atoms are deexcited. It is clear from the figure that before saturation $N_s$ is proportional to $P_L$ and $\chi^4$. As $\chi$ becomes large enough a significant amount of energy stored in the target is released and one gains very little by increasing the mixing or laser power. We note also that $N_\mathrm{rep}$, $N_\mathrm{pass}+1$ and $\sin^2(\mAp^2l/\omega)$ are will have the same scaling as $P_L$ when determining the number of signal photons emitted.

To estimate the sensitivity of our proposed experiment, we require the emission of at least ten photon pairs after a certain number of excitation/deexcitation repetitions. As a benchmark we take the laser power $P_L=1\ \mathrm{Wmm}^{-2}$, target area $A=1\ \mathrm{cm}^2$, target chamber length $L= 30$ am, number of dark photon-generating cavity reflections $N_\mathrm{pass}=2\times 10^4$, for cavity size $l=50\ \mathrm{cm}$, and number of experimental repetitions as $N_\mathrm{rep}=10^3$. In the regime that a fraction of the $\ph$ deexcites, the number of emitted photons can be estimated as
\begin{equation}
N_s=5 \times10^3\dfrac{P_L}{1\ \mathrm{Wmm}^{-2}}\left(\dfrac{\chi}{10^{-9}}\right)^4\dfrac{N_\mathrm{pass}+1}{2\times 10^4+1}\sin^2\left(1.27\dfrac{\mAp^2}{\mathrm{meV}^2}\dfrac{\mathrm{eV}}{\omega}\dfrac{l}{m}\right)\,,
\end{equation}
where this expression has been normalized assuming $n = 10^{21}~{\rm cm^{-3}}$.

We show the sensitivity of our proposed experiment in Figure~\ref{fig:sensitivity}. Also shown in the figure are the light-shining-through-wall experiments, cosmological, and astrophysical bounds reviewed in~\cite{Redondo:2015iea}. The coherent amplification condition we have assumed throughout given by Eq.~\eqref{eq:cohamp}, requires that our dark photon mass not be too large. This restricts $\mAp\lesssim 0.6\ \mathrm{meV}$. As a consequence we have truncated the mass sensitivity at one meV. As seen from the figure, over the mass range $10^{-5}\sim 10^{-3}\ $eV our proposed experiment appears rather sensitive to dark photon kinetic mixing. Note that till here we gave assumed the detuning $\delta=0$. A nonzero detuning has mild effect on the expected sensitivity. We refer the reader to Appendix~\ref{app:detuning} for quantitative discussion.

Looking at Figure~\ref{fig:sensitivity}, it is apparent that the number of signal photons depends on the number density of the target $n$ and the coherence factor $r_1$ in a nonlinear and nontrivial manner. To examine this behavior, we can combine Eqs.~\eqref{eq:eE1bloch}-\eqref{eq:eE'bloch} and for the moment neglect the propagation terms, $r_2$, and the position and time dependence of $r_1$ to obtain
\begin{equation}
(\partial^2_t-\partial^2_z)E_1-n^2\Omega_r^2E_1=0\,,
\label{eq:simpE1}
\end{equation}
where we define $\Omega_r^2 \equiv \omega^2|\aeg r_1|^2(1+\chi^2|\eta|^2)/4$. The temporal part of $E_1$ can be solved from Eq.~\eqref{eq:simpE1} which indicates $E_1\propto e^{n\Omega_rt}$. This gives the total photon yield 
\begin{equation}
N_s\propto \int|E_1|^2dt\sim \dfrac{1}{n\Omega}e^{2n\Omega_r\Delta t}\,,
\label{eq:signalapprox}
\end{equation}
with $\Delta t$ denoting the time duration of the experiment. The $\chi^2|\eta|^2$ term in $\Omega_r$ shows the dependence on the dark photon mixing parameter and relative polarization, since it is the dark photon that triggers the collective deexcitation of the target molecules. We should also keep in mind that coherence ($r_1$) dies off quickly after $\Delta t \sim T_2 \sim 10~{\rm ns}$, which causes the the drop in field intensity at $t\gtrsim T_2$ depicted in Figure~\ref{fig:fields_1e-3} and Figure~\ref{fig:fields_1e-9}. Therefore, Eq.~\eqref{eq:signalapprox} is no way close to an exact solution, but it does show that the signal intensity gets enhanced by a factor of $\sim e^{n\Omega_r t}$ even for a moderate increase in the number density $n$, coherence $r_1$, and the coherence time $T_2$. A related discussion can be found in~\cite{Yoshimura:2010jy}. This exponential evolution behavior can also be understood qualitatively. As mentioned in Section~\ref{sec:numerics}, the intial dark photon field triggers the emission of $E_1$ and $E_2$, which will subsequently trigger more two photon transitions. The number of photons to be triggered is proportional to the number density of the target, which appears in the exponent of the cascade deexcitation rate.

\section{Conclusions}
\label{sec:conc}
We have studied a new method to detect dark photon fields using resonant two photon de-excitation of coherently excited atoms. Our proposed experiment combines dark photon production techniques demonstrated by light-shining-through-wall experiments with a new detection method: dark photons triggering two photon transitions in a gas of parahydrogen coherently excited into its first vibrational state. The potential coupling sensitivity to dark photons we project in our benchmark setup is orders of magnitudes beyond present limits for ${\rm \mu eV - meV}$ mass dark photon fields.

A major technical hurdle to realizing this proposal will be the preparation of suitably coherent samples of cold parahydrogen using counter-propagating laser beams. As we examined in Section \ref{sec:2state}, the coherence times and $\ph$ densities necessary have already been achieved in laboratory conditions. It remains to suitably increase the fraction of coherently excited $\ph$ by using more powerful lasers and colder parahydrogen, as we explored in Section \ref{sec:2state}. However, even if complete parahydrogen sample coherence is not attained, it would still be possible to realize this proposal by increasing the density of parahydrogen, as explore in Section \ref{sec:sens}. Indeed, although we have not shown it in Figure \ref{fig:sensitivity}, the setup we propose with an increased $\ph$ number density ($2 \times 10^{21}$), assuming completely coherent atoms ($r_1 =1$) can probe kinetic mixings $\chi \ll 10^{-15}$. It may also be possible to realize a similar proposal to the one laid out here, using two photon nuclear transitions and free electron lasers. This might permit detecting dark photons at masses greater than an eV.

Our setup relies on the nonlinear development of electromagnetic fields in coherent atoms, and so our sensitivity estimates have relied on numerical simulations of dark photon and photon cascades in $\ph$. However as explained in Section \ref{sec:sens}, the proposed experiment will allow for dark photon detection to be directly calibrated using a low power trigger laser, as an equivalent stand-in for the dark photon field itself. While for this reason, we have focused on the detection of dark photons in this article, very similar methods could be used to detect axions and other light, electromagnetically-coupled particles. We leave this and other uses of multi stage atomic transitions to future work.

\section*{Acknowledgments}
We especially thank James Fraser for guidance on non-linear optics and laser power along with Junwu Huang, Sam McDermott, Alex Wright, and Aaron Vincent for useful discussions. Research at Perimeter Institute is supported by the Government of Canada through Industry Canada and by the Province of Ontario through the Ministry of Economic Development \& Innovation. A.~B., J.~B., and N.~S.~acknowledge the support of the Natural Sciences and Engineering Research Council of Canada.

\appendix

\section{Coherence and nonlinearity in two photon emission}
\label{sec:tpeqed}
Let us estimate the transition rate of two photon emission process, as illustrated in Figure~\ref{fig:transit_b}. The transition matrix  for
$|e\rangle\,\rightarrow\,|g\rangle$ transition is
\begin{align}
\langle g|iT|e\rangle&\simeq \mathcal{T}\dfrac{(-i)^2}{2}\int_{-\infty}^{+\infty}dt\int_{-\infty}^{+\infty}dt'\langle g |-\vec{d}\cdot\tilde{E}_2(t)|j\rangle \langle j| -\vec{d}\cdot\tilde{E}_1(t')|e\rangle\nonumber\\
&=(-i)^2\int_{-\infty}^{+\infty}dt\int_{-\infty}^{t}dt'\langle g |-\vec{d}\cdot\tilde{E}_2(t)|j\rangle \langle j| -\vec{d}\cdot\tilde{E}_1(t')|e\rangle\,.
\end{align}
where $\mathcal{T}$ is the time-ordering operator and we write the electric fields as
\begin{equation}
\tilde{E}_{m}=\dfrac{1}{2}E_m\vec{\epsilon} e^{-i\omega_m t+i\vec{k}\cdot\vec{r}}+\dfrac{1}{2}E^*_m\vec{\epsilon}^* e^{i\omega_m t-i\vec{k}\cdot\vec{r}}\,,\quad m=1,2\,,
\label{eq:Em}
\end{equation}
where $\omega_m$ and $\vec{k}_m$ are the energy and momentum of the emitted photons. Integrating over $t'$ yields
\begin{equation}
\langle g|iT|e\rangle \simeq i\dfrac{d_{je}d_{gj}}{\omega_1-\omega_{ej}}\dfrac{E_1E_2}{4}e^{-i(\vec{k}_1+\vec{k}_2-\vec{k}_{ej}^a)\cdot (\vec{r}-\vec{r}_a)}\int_{-\infty}^{+\infty}dt e^{i(\omega_1+\omega_2-\omega_{eg})t}\,,
\end{equation}
where as before we have defined $\omega_{ik}=\omega_i-\omega_k$ and $d_{ik}=\langle i|-\vec{d}\cdot \vec{\epsilon}^{(*)}|k \rangle$. $\vec{k}_{ej}^a$ is the change in the momentum of a specific $\ph$ after the transition and $r_a$ is the spatial position of the $\ph$. We can perform the second time integral and obtain
\begin{equation}
\langle g|iT|e\rangle=i2\pi\delta(\omega_{eg}-\omega_1-\omega_2)\mathcal{M}_a\,,
\end{equation}
with
\begin{equation}
\mathcal{M}_a=\dfrac{d_{gj}d_{je}}{\omega_{je}+\omega_1} \dfrac{E_1E_2}{4}e^{-i(\vec{k}_1+\vec{k}_2-\vec{k}_{eg}^a)\cdot (\vec{r}-\vec{r}_a)}=\dfrac{\aeg}{4}E_1E_2e^{-i(\vec{k}_1+\vec{k}_2-\vec{k}_{eg}^a)(\vec{r}-\vec{r}_a)}\,.
\end{equation}

First we consider the case that the $\ph$ is not emitting coherently, which we will call spontaneous two-photon deexcitation. In the case of spontaneous two-photon deexcitation, each $\ph$ emits two photons with frequencies that are not necessarily $\sim \weg/2$, in contrast with two photon emission induced by a trigger laser (where the trigger laser frequency used in earlier sections of this document matched the pump laser frequencies, all of these being $\weg/2$). In the spontaneous emission case, we sum up the contribution from all $\ph$ which gives the emission rate
\begin{equation}
\begin{split}
\Gamma_\mathrm{sp}=&\int\dfrac{d^3k_1}{(2\pi)^3}\dfrac{d^3k_2}{(2\pi)^3}\left|\int d^3 r\sum\limits_{a=1}^{N}\dfrac{\aeg}{4}\sqrt{\dfrac{4\omega_1\omega_2}{V^2}}e^{-i(\vec{k}_1+\vec{k}_2-\vec{k}_{eg}^a)(\vec{r}-\vec{r}_a)}\right|^2 \\
&\times 2\pi\delta(\omega_{eg}-\omega_1-\omega_2)\,,
\end{split}
\label{eq:Gammasp1}
\end{equation}
where we have explicitly replaced $E_m$ by $\sqrt{2\omega_m/V}$ for $m=1,2$, and $N$ is the number of spontaneous emitters. Since the exponential phase is random for each molecule, the product of the phases from different molecules will sum up to zero in the expansion of the square in Eq.~\eqref{eq:Gammasp1}. This gives
\begin{equation}
\Gamma_\mathrm{sp}=\int\dfrac{d^3k_1}{(2\pi)^3}\dfrac{d^3k_2}{(2\pi)^3}N|\aeg|^2\dfrac{\omega_1\omega_2}{4}2\pi\delta(\omega_{eg}-\omega_1-\omega_2)\,.
\end{equation}
Carrying out the integral we find
\begin{equation}
\dfrac{d\Gamma_\mathrm{sp}}{d\omega_1}=\dfrac{1}{(2\pi)^3}N|\aeg|^2\omega_1^3\omega_2^3\,.
\label{eq:dGammaspdw}
\end{equation}
If we define $z\equiv \omega_1/\weg$, Eq.~\eqref{eq:dGammaspdw} can be written as
\begin{equation}
\dfrac{d\Gamma_\mathrm{sp}}{dz}=\dfrac{\weg^7}{(2\pi)^3}N|\aeg|^2z^3(1-z)^3\,.
\label{eq:Gammasp2}
\end{equation}
We have used this equation to estimate the two photon spontaneous emission background in Sec.~\ref{sec:spont}.

Next we will estimate the rate for two photon emission for $\ph$ pumped and triggered in a manner which allows for macro superradiance. In the presence of appropriately applied background fields, $\ph$ molecules will tend to emit photons collectively with the same momenta. If the phase $k_{eg}^a$ is random for every molecule, the product of the phases would still cancel as we have derived before and the rate would be proportional to $N$; however, if the molecules are pumped into the excitation state {\it coherently} (by counter propagating lasers, in the setup we have considered), we can drop the superscript a in $k_{eg}^a$ and turn the sum into a spatial integral, $i.e.$
\begin{equation}
\begin{split}
\Gamma_\mathrm{sup}=&\int\dfrac{d^3k_1}{(2\pi)^3}\dfrac{d^3k_2}{(2\pi)^3}\left|\int d^3 r\int d^3r_a\dfrac{\aeg}{4}\rge n E_1E_2e^{-i(\vec{k}_1+\vec{k}_2-\vec{k}_{eg})(\vec{r}-\vec{r}_a)}\right|^2\\
& \times 2\pi\delta(\omega_{eg}-\omega_1-\omega_2)\,,
\end{split}
\end{equation}
where $n$ is the number density of the target and $\rge$ is the fraction of molecules in the coherent state. In the special case where we use two counter propagating lasers with the same frequency to pump the molecules, $\vec{k}_{eg} \approx 0$, although of course this can be spoiled by the lasers' linewidth and other experimental factors discussed in Section \ref{sec:2state}. For a dense and large enough target, the spatial integral in $r_a$ turns into a delta function, which gives
\begin{equation}
\Gamma_\mathrm{sup}=\int\dfrac{d^3k_1}{(2\pi)^3}\dfrac{d^3k_2}{(2\pi)^3}\left|\dfrac{\aeg}{4}\rge N E_1E_2 (2\pi)^3\delta^3(\vec{k}_1+\vec{k}_2-\vec{k}_{eg})\right|^2\times 2\pi\delta(\omega_{eg}-\omega_1-\omega_2)\,,
\end{equation}
where $N$ is the total number of $\ph$ in the target. In the case $\vec{k}_{eg} \approx 0$, the delta function forces $\vec{k}_1+\vec{k}_2=0$, meaning that the two photons emitted superradiantly have to be back-to-back and have equal frequency. Since the delta function is squared we replace one by the target volume $V$. Evaluating the integrals yields
\begin{equation}
\Gamma_\mathrm{sup}=\dfrac{1}{16\pi}|\aeg|^2|\rge|^2N^2V\omega_1^2|E_1|^2|E_2|^2\,.
\label{eq:Gammasup}
\end{equation}
Eq.~\eqref{eq:Gammasup} shows the transition rate in two photon superradiance is proportional to $N^2$ if coherence conditions are met. This can be compared with (out-of-phase) spontaneous two photon emission described in Eq.~\eqref{eq:Gammasp2} where the rate is proportional to $N$ instead. We also see that the rate grows nonlinearly with $E_1$ and $E_2$, the strength of the background fields. At the onset of superradiance, the emission rate is determined by the power of the trigger laser fields. As the photons from the deexcitation increase the strength of the electric fields, the deexcitation rate becomes larger and larger. This exponential growth is clearly seen in Figure~\ref{fig:fields_1e-3}.

\section{Estimate of dark photon triggered two stage transitions}
\label{app:base}

Let us now move on to estimate the emission rate of $\gamma_1$ and $\gamma_2$ in our proposed experiment, as depicted in Figure~\ref{fig:setup}. Consider the process illustrated in Figure~\ref{fig:transit_a}. First, the transition matrix for the deexcitation from $|e\rangle$ to $|g\rangle$ via the emission of a dark photon and a photon in the dark photon background is given by
\begin{align}
\langle g|iT|e\rangle&\simeq \mathcal{T}\dfrac{(-i)^2}{2}\int_{-\infty}^{+\infty}dt\int_{-\infty}^{+\infty}dt'\langle g |-\vec{d}\cdot\tilde{E'}(t)|j\rangle \langle j| -\vec{d}\cdot\tilde{E}_1(t')|e\rangle\nonumber\\
&=(-i)^2\int_{-\infty}^{+\infty}dt\int_{-\infty}^{t}dt'\langle g |-\vec{d}\cdot\tilde{E'}(t)|j\rangle \langle j| -\vec{d}\cdot\tilde{E}_1(t')|e\rangle\,,
\end{align}
where $\tilde{E}'$ and $E_1$ are given in a similar form as in Eq.~\eqref{eq:Em}.
 With same algebra as in Appendix~\ref{sec:tpeqed} we obtain
\begin{equation}
\langle g|iT|e\rangle=i2\pi\delta(\omega_{eg}-\omega_1-\omega')\mathcal{M}_a\,,
\end{equation}
with
\begin{equation}
\mathcal{M}_a=\dfrac{d'_{gj}d_{je}}{\omega_{je}+\omega_1} \dfrac{E'}{2}\sqrt{\dfrac{\omega_1}{2V}}e^{-i(\vec{k}_1+\vec{k}'-\vec{k}_{eg})\cdot (\vec{r}-\vec{r}_a})\,.
\end{equation}
Note that we have replaced $E_1$ by $\sqrt{2\omega_1/V}$ for one photon state. After introducing $\aeg$ and $\eta$ as defined in Eq.~\eqref{eq:age} and Eq.~\eqref{eq:eta}, we find
\begin{equation}
\mathcal{M}_a=\aeg \eta \dfrac{E'}{2}\sqrt{\dfrac{\omega_1}{2V}}e^{-i(\vec{k}_1+\vec{k}'-\vec{k}_{eg})\cdot \vec{r}_a}\,.
\label{eq:M}
\end{equation}
The transition rate, summing up all coherent atoms is now
\begin{equation}
\begin{split}
\Gamma_\mathrm{\gamma'\gamma}=&\int\dfrac{d^3k_1}{(2\pi)^3}\dfrac{d^3k'}{(2\pi)^3}\left|\int d^3 r\int d^3r_a\aeg\eta\rge n \dfrac{E'}{2}\sqrt{\dfrac{\omega_1}{2V}}e^{-i(\vec{k}_1+\vec{k}'-\vec{k}_{eg})\cdot(\vec{r}-\vec{r}_a)}\right|^2\\
& \times 2\pi\delta(\omega_{eg}-\omega_1-\omega')\,.
\end{split}
\end{equation}
After some algebra we arrive at the transition rate
\begin{equation}
\Gamma_\mathrm{\gamma'\gamma}=\dfrac{1}{8\pi}|\eta|^2|\aeg|^2|\rge|^2 N^2 \omega_1^3|E'|^2\,.
\end{equation}
With the dark photon field power given in Eq.~\eqref{eq:eprime0} we obtain
\begin{equation}
\Gamma_\mathrm{\gamma'\gamma}=\dfrac{1}{4\pi}(N_\mathrm{pass}+1)\chi^4\sin^2\left(\dfrac{\mAp^2}{4\omega'}l\right)P_L|\eta|^2|\aeg|^2|\rge|^2n^2V^2\omega_1^3\,.
\end{equation}
For $\chi=10^{-9}$, $\mAp=10^{-4}\ \mathrm{eV}$, $\omega'=\omega_1=0.26\ \mathrm{eV}$, $N_\mathrm{pass}=2\times 10^4$, $l=0.5\ \mathrm{m}$, $\eta=1$, $\ph$ number density $n=10^{21}\ \mathrm{cm}^{-3}$, target area $A=1\ \mathrm{cm}^2$, length $L=30\ \mathrm{cm}$, laser power $P_L=1$W/mm$^2$, $\aeg=0.0275\times10^{-24}\ \mathrm{cm}^3$,
\begin{equation}
\Gamma_\mathrm{\gamma'\gamma}=1.2\times 10^{-5}\ \mathrm{s}^{-1}\,.
\end{equation}
This emission rate is relatively low considering the experimental trial time of about $10\ \mathrm{ns}$, which is determined by the decoherence time. Signal photon production, on the other hand, is enhanced when the dark photon triggers two photon superradiant transitions. This is discussed in Sec.~\ref{sec:dpinducerate}. 

\section{Laser detuning}
\label{app:detuning}

\begin{figure}[!htb]
\centering
\begin{subfigure}{0.49\textwidth}
\includegraphics[width=\linewidth]{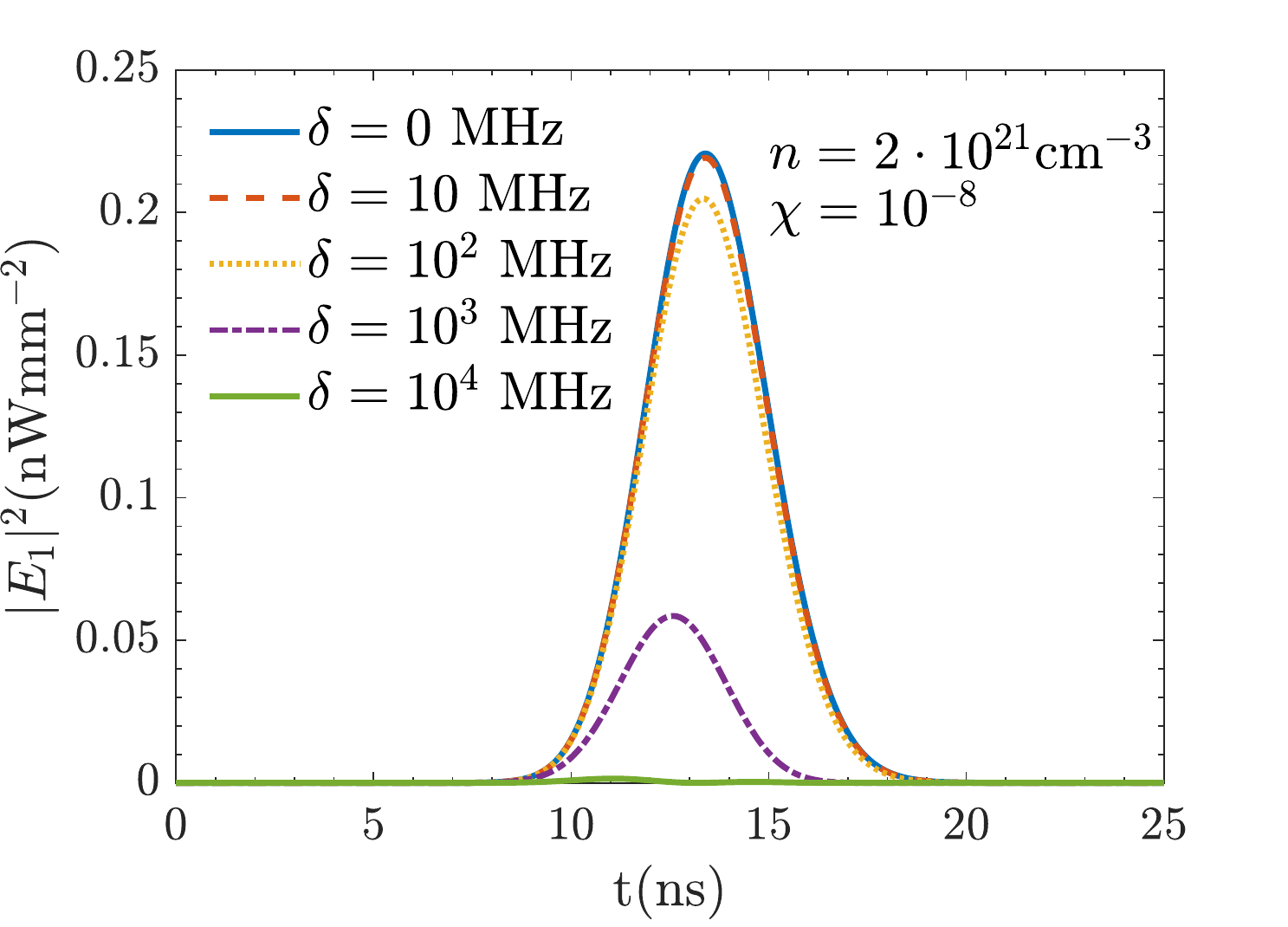}
\end{subfigure}
\hspace*{-0.15cm} 
\begin{subfigure}{0.49\textwidth}
\includegraphics[width=\linewidth]{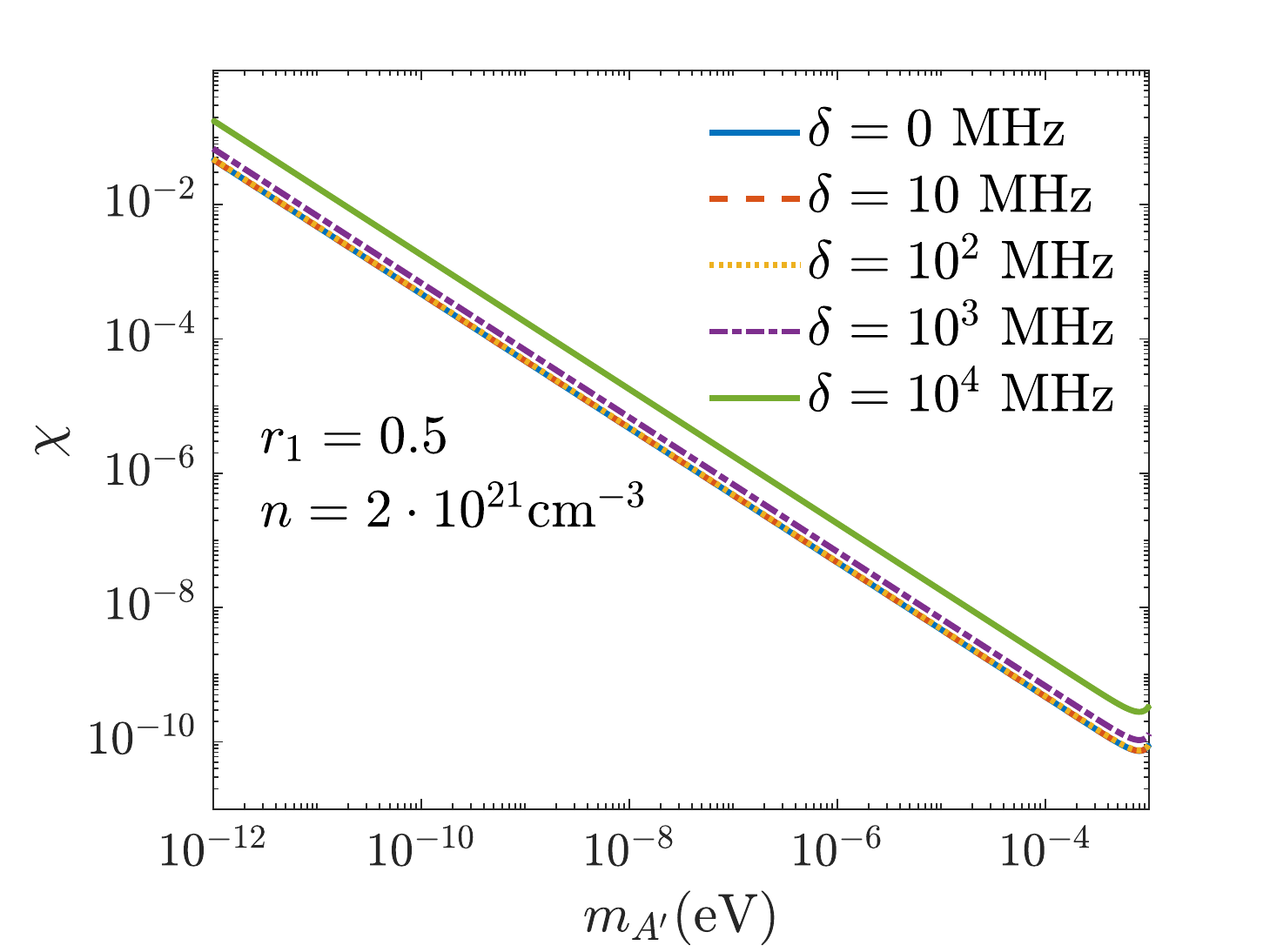}
\end{subfigure}
\caption{Effect of varying the detuning parameter. Left panel: Time evolution of the electric field at the end of the $\ph$ target for $\delta=0$ (solid blue), $\delta=10$~MHz (dashed red), $\delta=100$~MHz (dotted yellow), $\delta=1000$~MHz (dash-dotted purple), $\delta=10000$~MHz (solid green). We assume the benchmark setup given in Table \ref{tab:bench}, except for the $\ph$ number density $n=2\times 10^{21}$~cm$^{-3}$. We have used initial coherence values $r_1(t=0)=0.5$, $r_2(t=0)=r_3(t=0)=0$, dark photon mixing parameter $\chi=10^{-8}$, and dark photon mass $\mAp=0.1$~meV. Note that the electric field power is given in units of 
$10^{-9}$~Wmm$^{-2}$. Right panel: Projected sensitivity of our proposed experiment for different detunings. The experimental setup is the same as for the left panel.}
\label{fig:deltaeff}
\end{figure}

In this appendix we  study how the development of signal photons are altered if the counter-propagating lasers used to excite the cold parahydrogen sample are substantially detuned. This amounts to varying the detuning parameter $\delta$. The primary effect of $\delta$ is to induce the oscillations in the coherence factor $\rge$ (hence also the Bloch vectors $r_1$ and $r_2$), as indicated by Eq.~\eqref{eq:erge}. We show the effect of $\delta$ in Fig.~\ref{fig:deltaeff}. As $\delta$ increases, the coherence of $\ph$ is suppressed, and the output photon flux when $\ph$ de-excites is correspondingly suppressed. However, as shown there is no suppression for $\delta<100$~MHz. We note that detuning $\delta<100$~MHz has already been achieved in existing experiments \cite{Hiraki:2018jwu}. After increasing the detuning to a value as large as $\delta=10^4$~MHz there is a notable decrease in the expected experimental sensitivity, since the resulting $\ph$ sample will be less coherent. We note that, as before, the experimental sensitivity and $\ph$ coherence are obtained by numerically solving the differential equations given in Eqs.~\eqref{eq:er1} through \eqref{eq:eE'bloch}, which determines the nonlinear development of coherence in $\ph$.

\bibliography{atomictransition}
\bibliographystyle{JHEP}


%
%
%

\end{document}